\documentclass[12pt,preprint]{aastex}

\shorttitle{Transverse oscillations of systems of coronal loops}
\shortauthors{M. Luna et al.}
\begin{document}

\title{Transverse oscillations of systems of coronal loops}

\author{M. Luna\altaffilmark{1}, J.
Terradas\altaffilmark{2}, R. Oliver\altaffilmark{1}, and J.L.
Ballester\altaffilmark{1}}

\altaffiltext{1}{Departament de F\'{\i}sica, Universitat de les Illes Balears,
07122 Palma de Mallorca, Spain. Email: manuel.luna@uib.es,
jaume.terradas@uib.es, ramon.oliver@uib.es and joseluis.ballester@uib.es}
\altaffiltext{2}{Centre for Plasma Astrophysics, Katholieke Universiteit Leuven,
Celestijnenlaan 200B, B-3001 Leuven, Belgium}

\begin{abstract}

We study the collective kinklike normal modes of a system of several cylindrical
loops using the T-matrix theory. Loops that have similar kink frequencies
oscillate collectively with a frequency which is slightly different from that of
the individual kink mode. On the other hand, if the kink frequency of a loop is
different from that of the others, it oscillates individually with its own
frequency. Since the individual kink frequency depends on the loop density but
not on its radius for typical $1$ MK coronal loops, a coupling between kink
oscillations of neighboring loops take place when they have similar densities.
The relevance of these results in the interpretation of the oscillations studied
by \citet{schrijver2000} and \citet{verwichte2004}, in which transverse
collective loop oscillations seem to be detected, is discussed. In the first
case, two loops oscillating in antiphase are observed; interpreting this motion
as a collective kink mode suggests that their densities are roughly equal. In
the second case, there are almost three groups of tubes that oscillate with
similar periods and therefore their dynamics can be collective, which again
seems to indicate that the loops of each group share a similar density. All the
other loops seem to oscillate individually and their densities can be different
from the rest.

\end{abstract}


\section{Introduction}

Transverse coronal loop oscillations were discovered by the Transition Region
and Coronal Explorer (TRACE) in 1998 \citep[see,
e.g.][]{aschwanden1999,aschwanden2002, nakariakov1999}. These oscillations were
initiated shortly after a solar flare that disturbed the loops. Since their
first observation, transverse oscillations have been routinely observed and
studied. Much before TRACE observations, the theory of loop oscillations was
developed \citep{spruit1981,edwin&roberts1983,cally1986} and the different kinds
of oscillations were studied. The observed transverse motions have been
interpreted in terms of the fundamental kink mode of the fast
magnetohydrodynamic (MHD) oscillation \citep{nakariakov1999}, which is the only
mode that can produce the observed transverse loop displacement.

In many cases the observed coronal loops belong to complex active regions and
are not isolated but forming bundles or arcades of loops. For example, in
\citet{schrijver2000} antiphase transverse oscillations of adjacent loops were
reported. In addition, in \citet{verwichte2004} phase and antiphase motions were
observed in a post-flare arcade. On the other hand, it is currently debated
whether active region coronal loops are monolithic or multistranded \citep[see,
e.g.][]{aschwanden2005, klimchuk2006,deforest2007}. In the multistranded model,
it is suggested that loops are formed by several tens or hundreds of strands
considered as miniloops for which the heating plasma properties are
approximately uniform in the transverse direction \citep{klimchuk2006}. Most
analytical studies about transverse loop oscillations have only considered the
properties of individual loops. However, from the information provided by the
observations, it is necessary to study not only individual loops but also how
several tubes can oscillate as a whole, since their joint dynamics can be
different from that of a single loop. Only a few works have considered composite
structures. \citet{berton1987} studied the MHD normal modes of a periodic
magnetic medium. \citet{kris93} and \citet{kris94} studied numerically the
propagation of fast waves in two slabs unbounded in the longitudinal direction.
In \citet{diaz2005} the oscillations of the prominence thread structure were
investigated. These authors found that in a system of equal fibrils the only
non-leaky mode is the symmetric one, which means that all the fibrils oscillate
in spatial phase with the same frequency.  \citet{luna2006} studied a system of
two coronal slabs and found that the symmetric and antisymmetric modes can be
trapped. A more complex system of two coronal cylinders was studied in
\citet{luna2008}. Four trapped normal modes were found and the interchange of
energy between loops was shown by solving the time-dependent problem. In
\citet{terradas2008} a multistranded loop formed by ten strands was considered.
The composite loop oscillates transversely as a whole with a global motion of
the strands after an external disturbance. This work shows that the bundle of
strands oscillates with a combination of collective modes. On the other hand, an
analytical approximation to the normal modes of a loop pair has been carried out
by \citet{doorsselaere2008}. The authors assume the long wavelength
approximation and obtain an analytical dispersion relation for two different
tubes together with the four kink mode polarizations described in
\citet{luna2008}.

In this work we aim to study the normal modes of a loop set with different
physical and geometrical properties by using the scattering theory. The
scattering theory, or its matricial formulation called T-matrix theory
\citep[see, e.g.][]{waterman&truell1961,waterman1969,ramm}, was first applied to
magnetic tubes by \citet{bogdan&zweibel1985}. These authors studied the
interaction of acoustic plane waves with an ensemble of parallel magnetic
fibrils distributed uniformly in the so-called spaghetti sunspot model. The
authors derived and solved the dispersion relation in the long wavelength limit.
In \cite{bogdan&cattaneo1989} the frequency shifts and velocity eigenfunctions
were calculated for the case of random fibril distributions of up to 100 flux
tubes. Many other papers were published studying the cross section of a fibril
spot insonified by external acoustic waves
\cite[see][]{bogdan&fox1991,keppens1994}. In all these papers a non-magnetized
external medium was considered and the eigenfrequencies and eigenmodes of the
acoustic oscillations were obtained.

In this paper we generalize the method to a system with an external magnetized
medium, in order to extend previous works to coronal loop conditions. Our model
consists of an ensemble of parallel cylinders, without gravity and curvature. We
consider uniform magnetic field in the internal loop medium and in the external
or coronal medium. This assumption allows the interaction of the tubes through
fast MHD waves. In addition, we explicitly calculate the eigenvalues and
eigenfunctions of the normal modes of the model.

This paper is organized as follows. In \S \ref{model} the loop ensemble model
and the equations for its dynamics are presented. In \S \ref{normal_modes} we
briefly describe the T-matrix theory and apply it to our model. With this
theory the exact eigenfrequencies and eigenmodes of two non-identical loops are
investigated in \S \ref{2loops}. We study the dependence of the interaction with
the relative density and radii of the loops. The study of three identical
aligned loops is presented in \S \ref{3loops}. In the same section the
interaction between three non-identical loops is considered. Finally in \S
\ref{disc_conc} the results are summarized and the main conclusions are drawn.

\section{Theoretical model}\label{model}

The equilibrium configuration used to model the loop set is a system of
$N$-cylindrical, parallel homogeneous flux tubes, with the $z$-axis pointing in
the direction of the loop axes. All loops have the same length, $L$, and each
individual loop, labeled $j$, is characterized by the position of its center in
the $xy$-plane, $\mathbf{r}_\mathrm{j}= x_\mathrm{j} \mathbf{e}_x + y_\mathrm{j}
\mathbf{e}_y$, its radius, $a_\mathrm{j}$, and its density, $\rho_\mathrm{j}$.
The density of the coronal environment is $\rho_\mathrm{0}$. The tubes and the
external medium are permeated by a uniform magnetic field along the
$z$-direction ($\mathbf{B}_0=B_0 \mathbf{e}_\mathrm{z}$). The Alfv\'en speed,
$v_\mathrm{A}=B_0/\sqrt{\mu \rho}$, takes the value $v_\mathrm{A j}$ inside the
$j$-th loop and $v_\mathrm{A 0}$ in the surrounding corona ($v_\mathrm{A
j}<v_\mathrm{A 0}$).

Linear perturbations about this equilibrium for a perfectly conducting fluid can
be readily described using the ideal MHD equations. In the zero-$\beta$ limit
these equations can be written as
\begin{equation}\label{wave_eq_pt}
 \left(\frac{\partial^2}{\partial t^2}-v_\mathrm{A}^2 \nabla^2\right) p_\mathrm{T}=0 ,
\end{equation}
where $p_\mathrm{T}$ is the total pressure perturbation
\begin{equation}\label{def_pres_pert}
p_\mathrm{T}=\frac{B_0}{\mu} B_z ,
\end{equation}
and $B_z$ is the $z$-component of the magnetic field perturbation. The other
perturbed quantities, namely the velocity, $\mathbf{v}$, the magnetic field
perturbation, $\mathbf{B}$, and the density perturbation, $\mathbf{\rho}$, can
be derived from $p_\mathrm{T}$. We have assumed a $z$-dependence of the
perturbations of the form $e^{-i k_z z}$. The line-tying effect is incorporated
by setting $k_z=q \pi /L$ where $q$ is an integer number. Hereafter we
concentrate on the fundamental mode and take $q=1$. We only consider problems
for which the time dependence is a simple harmonic oscillation with frequency
$\omega$. Then, the total pressure perturbation can be written in cylindrical
coordinates as
\begin{equation}
 p_\mathrm{T} = e^{i(k_z z -\omega t)} \psi(r,\varphi) ,
\end{equation}
where $\psi(r,\varphi)$ is a function that includes the dependence on $r$ and
$\varphi$. Inserting this expression in equation (\ref{wave_eq_pt}), we obtain
the scalar Helmholtz equation
\begin{equation}\label{helm_eq}
 \nabla_\perp^2 \psi + k_\perp^2 \psi=0 ,
\end{equation}
where $\perp$ refers to the direction perpendicular to the magnetic field
$\mathbf{B}_0$, i.e. to the $z$-axis, and $k_\perp$ is
\begin{equation}\label{k_def}
 k_\perp^2=\frac{\omega^2-k_z^2 v_\mathrm{A}^2}{v_\mathrm{A}^2} .
\end{equation}
Hereafter, the $\perp$ symbol is dropped for the sake of simplicity.

\section{Normal modes}\label{normal_modes}

The scattering theory, or its matricial formulation called T-matrix theory,
provides an scheme to find analytically the normal modes of a system of
scatterers in which waves are described by a Helmholtz equation \citep{ramm}. We
fulfill the T-matrix theory requirements because our ensemble of $N$ loops can
be considered a collection of scatterers and the perturbed total pressure is
described by equation (\ref{helm_eq}).

The T-matrix scheme states that the $j$-th flux tube generates an outgoing
scattered wave, $\psi_\mathrm{sc}^\mathrm{j}$, in a field position $\mathbf{r}$
(in the two-dimensional  $xy$-plane) that adds to the waves scattered from the
other loops to produce the following net external field
\citep{bogdan&cattaneo1989}
\begin{equation}\label{net_field}
\psi(\mathbf{r})=\sum_\mathrm{j}^\mathrm{N}\psi_\mathrm{sc}^\mathrm{j}(\mathbf{r}-\mathbf{r}_\mathrm{j})
.
\end{equation}
The scattered wave by the $j$-th loop is produced as a response of an exciting
wave produced by the external field minus the contribution of the mentioned
loop,
\begin{equation}\label{excitation_field}
\psi_\mathrm{ex}^\mathrm{j}(\mathbf{r}-\mathbf{r}_\mathrm{j})=\psi(\mathbf{r})-\psi_\mathrm{sc}^\mathrm{j}(\mathbf{r}-\mathbf{r}_\mathrm{j})
.
\end{equation}
With equations (\ref{net_field}) and (\ref{excitation_field}) the exciting
field, $\psi_\mathrm{ex}^\mathrm{j}$, may be written entirely in terms of the
scattered field, resulting in the self-consistency field equation
\citep{bogdan&cattaneo1989}. This system of equations may then be closed by
noting that the exciting and scattered fields are further related by linear
operators, $\mathbf{T}^\mathrm{j}$, that describe the scattering properties of
the individual flux tubes \citep{bogdan&cattaneo1989,waterman1969,ramm}
\begin{equation}\label{Tmatrix_relation}
\psi_\mathrm{sc}^\mathrm{j} (\mathbf{r}-\mathbf{r}_\mathrm{j})=
\mathbf{T}^\mathrm{j} \psi_\mathrm{ex}^\mathrm{j}
(\mathbf{r}-\mathbf{r}_\mathrm{j}).
\end{equation}
The key point is that the linear operators $\mathbf{T}^\mathrm{j}$ depend
exclusively on the individual loop and external medium properties and can be
directly computed through the boundary conditions on the loop-external medium
interphase as we will see below.

The external field to the $j$-th loop can be decomposed with equation
(\ref{excitation_field}) as an excitation field on this loop and a scattered
field by this loop. The excitation field has no sources in the $j$-th loop, i.e.
it is the scattered field of the other loops, so it can be written as
\begin{equation}\label{excitation_field_explicit}
\psi_\mathrm{ex}^\mathrm{j}
(R_\mathrm{j},\varphi_\mathrm{j})=\sum_{\mathrm{m}=-\infty}^{\infty}
\alpha_\mathrm{m}^\mathrm{j} J_\mathrm{m} (k_\mathrm{0} R_\mathrm{j}) e^{i
\mathrm{m} \varphi_\mathrm{j}} ,
\end{equation}
where $\alpha_\mathrm{m}^\mathrm{j}$ are the expansion coefficients of order
$m$, that depend on the $k_z$ wave number and the frequency $\omega$, and
$R_\mathrm{j}$ and $\varphi_\mathrm{j}$ are the local polar coordinates centered
at $\mathbf{r}_\mathrm{j}$, defined through
$R_\mathrm{j}=|\mathbf{r}-\mathbf{r}_\mathrm{j}|$ and $\cos \varphi_\mathrm{j}
=\mathbf{e}_x \cdot
(\mathbf{r}-\mathbf{r}_\mathrm{j})/|\mathbf{r}-\mathbf{r}_\mathrm{j}|$. Here
$J_\mathrm{m}$ is the Bessel function of the first kind and order $m$ and
$k_\mathrm{0}$ is $k$ in the external medium calculated using equation
(\ref{k_def}). With equations (\ref{Tmatrix_relation}) and
(\ref{excitation_field_explicit}), we find the scattered field in terms of an
outgoing wave with sources at $\mathbf{r}_\mathrm{j}$,
\begin{equation}\label{scattered_field_explicit}
 \psi_\mathrm{sc}^\mathrm{j}
	(R_\mathrm{j},\varphi_\mathrm{j})=\sum_{\mathrm{m}=-\infty}^{\infty}
	T_\mathrm{m m}^\mathrm{j} \alpha_\mathrm{m}^\mathrm{j}
	H^{(\mathrm{1})}_\mathrm{m} (k_\mathrm{0} R_\mathrm{j}) e^{i \mathrm{m}
	\varphi_\mathrm{j}} ,
\end{equation}
where $T_\mathrm{m m}^\mathrm{j}$ are the matrix diagonal elements of the
operator $\mathbf{T}^\mathrm{j}$ projected on the local basis, called T-matrix.
The non-diagonal elements of this matrix are zero for axisymmetric tubes
\citep{bogdan&zweibel1985}. The functions $H^{(\mathrm{1})}_\mathrm{m}$ are the
Hankel functions of the first kind. With equations (\ref{net_field}),
(\ref{excitation_field}), (\ref{excitation_field_explicit}), and
(\ref{scattered_field_explicit}), we find the following expression for the total
field
\begin{equation}\label{netfield_ex+sc}
 \psi(\mathbf{r})=\sum_{\mathrm{m}=-\infty}^{\infty}
	\alpha_\mathrm{m}^\mathrm{j} \left[  J_\mathrm{m} (k_\mathrm{0}
	R_\mathrm{j}) + T_\mathrm{m m}^\mathrm{j}  H^{(\mathrm{1})}_\mathrm{m}
	(k_\mathrm{0} R_\mathrm{j}) \right] e^{i \mathrm{m} \varphi_\mathrm{j}},
\end{equation}
in which the external field to $j$-th loop is decomposed as an excitation on
this loop and a scattered field by this loop \citep{waterman1969}.

Following the development of \citet{bogdan&cattaneo1989}, a linear algebraic
system of equations for the complex coefficients $\alpha_\mathrm{m}^\mathrm{j}$
may then be obtained. We first substitute equation (\ref{net_field}) in equation
(\ref{excitation_field}) in order to obtain the self-consistency requirement
\begin{equation}\label{self_consistency}
\psi_\mathrm{ex}^\mathrm{j}(\mathbf{r}-\mathbf{r}_\mathrm{j})=\sum_{\mathrm{i} \neq \mathrm{j}}^{\mathrm{N}} \psi_\mathrm{sc}^\mathrm{i}(\mathbf{r}-\mathbf{r}_\mathrm{i}) .
\end{equation}
Next the exciting and scattered fields are replaced by their basis expansions, equations
(\ref{excitation_field_explicit}) and (\ref{scattered_field_explicit}), and the translation formula \citep[see appendix of][]{bogdan&cattaneo1989} is used to
express the scattered wave centered in the $i$-th loop into an excitation
at $j$-th flux tube. Finally, we obtain the following set of equations
\begin{equation}\label{problem_solution}
\alpha_\mathrm{m}^\mathrm{j}+\sum_ {\mathrm{i} \neq \mathrm{j}}^{\mathrm{N}}\sum_{\mathrm{n}=-\infty}^{\infty} \alpha_\mathrm{n}^\mathrm{i} T_\mathrm{n n}^\mathrm{i}
H_\mathrm{n-m}^{(\mathrm{1})}(\mathrm{k}_\mathrm{0}|\mathbf{r}_\mathrm{j}-\mathbf{r}_\mathrm{i}|) e^{i (\mathrm{n}-\mathrm{m}) \varphi_\mathrm{j i}} =0 ,
\end{equation}
where $\varphi_\mathrm{j }$ is the angle formed by the center of the $i$-th loop
with respect to the center of the $j$-th flux tube. As we will see below with
this equation we can find the $\alpha_\mathrm{m}^\mathrm{j}$ coefficients and
the frequencies $\omega$,  from which the spatial structure of the normal modes
can be determined. From equation (\ref{problem_solution}), we see that the
expansion coefficient of order $m$ of the $j$-th loop,
$\alpha_\mathrm{m}^\mathrm{j}$, is coupled to all expansion coefficients of the
other loops. This fact reflects the collective nature of the normal modes. With
the $\alpha_\mathrm{m}^\mathrm{j}$ and equations (\ref{net_field}) and
(\ref{scattered_field_explicit}) we find the net external field. 

The internal or transmitted field is
\begin{equation}\label{tansmited_field_explicit}
\psi_\mathrm{tr}^\mathrm{j}(\mathbf{r}-\mathbf{r}_\mathrm{j})=\sum_{\mathrm{m}=-\infty}^{\infty} 
	\beta_\mathrm{m}^\mathrm{j} J_\mathrm{m} (k_\mathrm{j} R_\mathrm{j})
	e^{i \mathrm{m} \varphi_\mathrm{j}} ,
\end{equation}
where $k_\mathrm{j}$ is the transverse wavenumber inside the $j$-th loop
calculated using equation (\ref{k_def}). The Bessel functions of the second
kind, $Y_\mathrm{m}$, are not considered in the expansion of the internal field
(eq.\~[\ref{tansmited_field_explicit}]) because they are singular at the loop
axes.  The transmitted field (eq.\~[\ref{tansmited_field_explicit}]) can be
calculated through the boundary conditions, namely the continuity of the total
pressure perturbation (\ref{def_pres_pert}) and the radial component of the
velocity at $R_\mathrm{j}=a_\mathrm{j}$ \citep[see][]{goedbloed1983}. In terms
of the $\psi$ fields they are expressed as follows
\begin{eqnarray}\label{first_bc}
\psi_\mathrm{tr}^\mathrm{j} (k_\mathrm{j}
R_\mathrm{j})|_{R_\mathrm{j}=a_\mathrm{j}}&=&\psi (k_\mathrm{0}
R_\mathrm{j})|_{R_\mathrm{j}=a_\mathrm{j}} , \\ \label{second_bc}
\frac{{\psi_\mathrm{tr}^\mathrm{j}}'(k_\mathrm{j}
R_\mathrm{j})|_{R_\mathrm{j}=a_\mathrm{j}}}{k_\mathrm{j}}&=&\frac{ \psi'
(k_\mathrm{0} R_\mathrm{j})|_{R_\mathrm{j}=a_\mathrm{j}}}{k_\mathrm{0}} ,
\end{eqnarray}
where the prime is the derivative with respect to the function argument, $\psi'
(x) = \partial \psi (x) / \partial x$.

Equation (\ref{problem_solution}) is completely general for a system of $N$
cylindrical flux tubes \citep[see][for non-axisymmetric
expressions]{keppens1995} and all the information of the individual loops is
included in the T-matrix elements, $T_\mathrm{m m}^\mathrm{j}$. These elements
are calculated through the boundary conditions at the interphase between the
loop and the external medium.  With equations (\ref{netfield_ex+sc}),
(\ref{tansmited_field_explicit}), (\ref{first_bc}), and (\ref{second_bc}) we
find the $T_\mathrm{m m}^\mathrm{j}$ element expression
\begin{equation}\label{tmatrix_explicit}
 T^\mathrm{j}_\mathrm{m m}=\frac{{k_\mathrm{j}}^2 k_\mathrm{0} J_\mathrm{m}
	(k_\mathrm{j} a_\mathrm{j}) {J'}_\mathrm{m} (k_\mathrm{0}
	a_\mathrm{j})-{k_\mathrm{0}}^2 k_\mathrm{j} {J'}_\mathrm{m}
	(k_\mathrm{j} a_\mathrm{j}) {J}_\mathrm{m} (k_\mathrm{0}
	a_\mathrm{j})}{{k_\mathrm{0}}^2 k_\mathrm{j}
	{H}_\mathrm{m}^{(\mathrm{1})}(k_\mathrm{0} a_\mathrm{j}) {J'}_\mathrm{m}
	(k_\mathrm{j} a_\mathrm{j})-{k_\mathrm{j}}^2 k_\mathrm{0}
	{H'}_\mathrm{m}^{(\mathrm{1})}(k_\mathrm{0} a_\mathrm{j}) J_\mathrm{m}
	(k_\mathrm{j} a_\mathrm{j})} .
\end{equation}
Equation (\ref{tmatrix_explicit}) is the generalization of
\citet{bogdan&zweibel1987,bogdan&cattaneo1989}  to the case of a magnetized
environment. The zeroes of the denominator correspond to the dispersion relation
of the individual loop \cite[see, e.g.][]{cally1986}.

Finally, note that with the boundary conditions (eqs.\~[\ref{first_bc}] and
[\ref{second_bc}]) it is possible to calculate the $\beta_\mathrm{m}^\mathrm{j}$
coefficients
\begin{equation}
 \beta_\mathrm{m}^\mathrm{j}=\frac{J_\mathrm{m}(k_\mathrm{0} a_\mathrm{j})+T^\mathrm{j}_\mathrm{m
	m} H_\mathrm{m}^{(\mathrm{1})} (k_\mathrm{0} a_\mathrm{j}) }{J_\mathrm{m} (k_\mathrm{j} a_\mathrm{j})} \alpha_\mathrm{m}^\mathrm{j},
\end{equation}
which can be inserted into equation (\ref{tansmited_field_explicit}) to obtain
the internal field, $\psi_\mathrm{tr}^\mathrm{j}$.

Equation (\ref{problem_solution}) is formally an infinite system of equations
for an infinite number of unknowns ($\alpha_\mathrm{m}^\mathrm{j}$). In order to
solve it, we truncate the system into a finite number of equations and unknowns
by setting $\alpha^\mathrm{j}_{\mathrm{m_\mathrm{t}}+1}=0$ for azimuthal numbers
greater than a truncation number ($m>m_\mathrm{t}$). To ensure the convergence
of solutions, they must be independent of the truncation number $m_\mathrm{t}$.
With these considerations, the solution of equation (\ref{problem_solution})
reduces to solve a homogeneous linear system of $N (2 m_\mathrm{t}+1)$ equations
and $N (2 m_\mathrm{t}+1)$ unknowns. For this system of equations to have a
non-trivial solution, its determinant must be zero. This requirement gives the
dispersion relation as a transcendent equation. We solve the dispersion relation
numerically and find the frequencies of the normal modes and with these
frequencies we calculate the $\alpha^\mathrm{j}_\mathrm{m}$ expansion
coefficients. With equations (\ref{net_field}) and
(\ref{tansmited_field_explicit}) we find the net field in the external medium
and the transmitted field in each loop. In all our calculations, solutions are
independent of the truncation number for values $m_\mathrm{t}>5$ but we fix this
number to $m_\mathrm{t}=20$ to more confidently ensure their convergence. With
the method presented here we have obtained the results of the following
sections.


\section{Interaction between two loops}\label{2loops}

\begin{figure}[!ht]
\center
\resizebox{11.5cm}{!}{\includegraphics{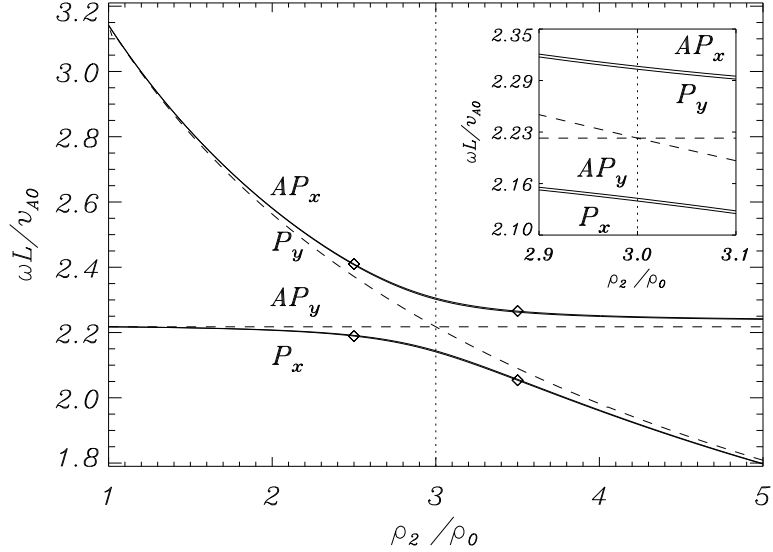}}
\caption{
Dimensionless frequency, $\omega L/v_\mathrm{A 0}$, as a function of the
internal density of the second loop. The bottom solid line is associated to the
two kinklike normal modes $P_x$ and $AP_y$, which have very similar frequencies.
In the same way, the upper solid line is associated to the $P_y$ and $AP_x$
modes. In the inner plot a detailed view for $\rho_\mathrm{2} \approx
\rho_\mathrm{1}$ is displayed showing that the solid lines are in fact double
lines. The two dashed lines are the individual kink frequencies of each loop. We
see that the kink frequency of loop $1$ is constant and the frequency of loop
$2$ decreases with $\rho_\mathrm{2}$, because $\rho_\mathrm{1}$ is constant and
$\rho_\mathrm{2}$ changes. The vertical dotted line is plotted at
$\rho_\mathrm{2}=\rho_\mathrm{1}$. Diamonds mark the frequencies of the modes
represented in Fig. \ref{2dif_loops}.}
\label{ratio_densities}
\end{figure}

First, we compute the normal modes of two non-identical loops with the T-matrix
theory outlined in \S \ref{normal_modes}. In this section we study the
dependence of the interaction as a function of the density and radii of the
loops. We consider a system of two loops with radii
$a_\mathrm{1}=a_\mathrm{2}=a=0.03 L$ and separated a distance $d=3 a$. The first
loop density is $\rho_\mathrm{1}=3 \rho_\mathrm{0}$ while $\rho_\mathrm{2}$ is
changed from $\rho_\mathrm{2}=\rho_\mathrm{0}$ to $5 \rho_\mathrm{0}$ to study
its influence on the normal mode properties. We concentrate on the kinklike
modes in which the individual loops move more or less as kink and suffer
the largest transverse displacement. There are other higher order normal modes
whose spatial structure is more complex, i.e. fluting modes. We find four
kinklike normal modes named $P_x$, $AP_y$, $P_y$, and $AP_x$, where $P$ and $AP$
refer to phase or antiphase motions of the loops, respectively, and the
subscript $x$ or $y$ refers to the direction of the motion along the $x$- or
$y$-axes. The frequencies of oscillation of these four modes as a function of
$\rho_\mathrm{2}/\rho_\mathrm{0}$ are displayed in Figure \ref{ratio_densities}.
The bottom solid line is associated to the $P_x$ and $AP_y$ modes, which almost
have the same frequency (see inbox Fig.~\ref{ratio_densities}). The same
behavior is found for the top solid line, which corresponds to the $P_y$ and
$AP_x$ modes. We see that the collective normal modes (solid lines) do not
coincide with the kink frequencies of the individual loops (dashed lines), a
discrepancy caused by the interaction between loops. This interaction is maximal
when $\rho_\mathrm{2} = \rho_\mathrm{1}$ (dotted line), and the normal modes
$P_x$, $AP_y$, $P_y$, and $AP_x$ become the modes $S_x$, $A_y$, $S_y$, and $A_y$
reported in \citet{luna2008}. The opposite situation takes place when
$\rho_\mathrm{2}$ is sufficiently different from $\rho_\mathrm{1}$: the
collective frequencies are closer to the individual kink frequencies and the
system behaves as a pair of independently oscillating loops. In this regime, the
$P_x$ and $AP_y$ modes correspond to the individual oscillations of the denser
loop in the $x$- and $y$-direction respectively and possess identical
frequencies, whereas the $P_y$ and $AP_x$ modes are the individual oscillations
of the rarer loop in the $x$- and $y$-direction, respectively, and also share
the same frequency. Figure \ref{ratio_densities} can be interpreted globally as
an avoided crossing of the kink modes of the loops: far from the coupling, each
branch is associated to the individual loop kink mode, but near the avoided
crossing motions are associated to the two loops to produce four collective
kinklike modes. As for as kinklike solutions are concerned, loops interact for
densities approximately in the range $\rho_\mathrm{2}=2 \rho_\mathrm{0}$ to $4
\rho_\mathrm{0}$.

\begin{figure}[!ht]
\center
\hspace{-1.5cm}
\mbox{\hspace{-1.cm}\hspace{8.5cm}\hspace{8.5cm}\hspace{2.cm}}
\includegraphics[width=8.5cm]{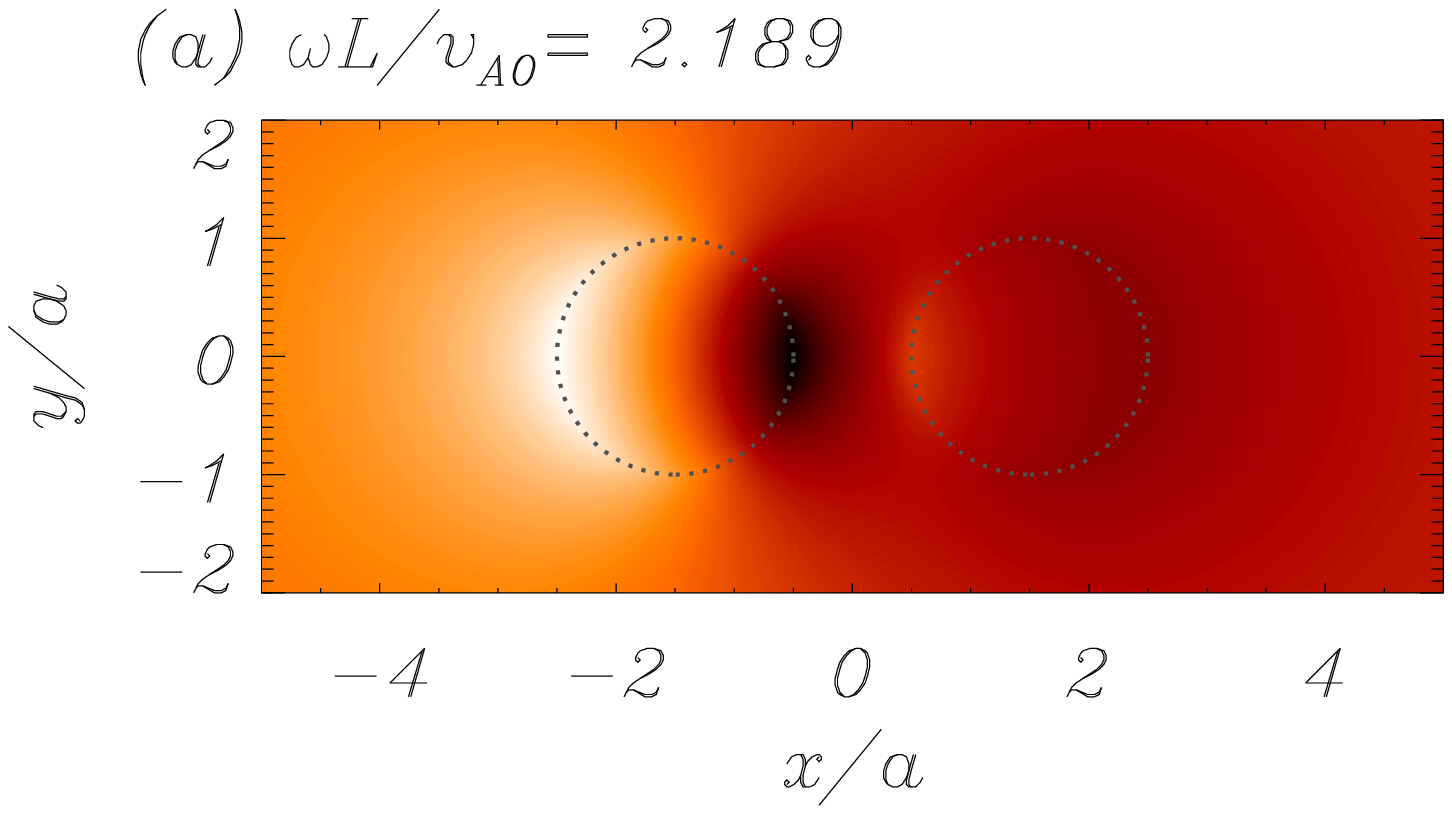}\hspace{-1.3cm}\includegraphics[width=8.5cm]{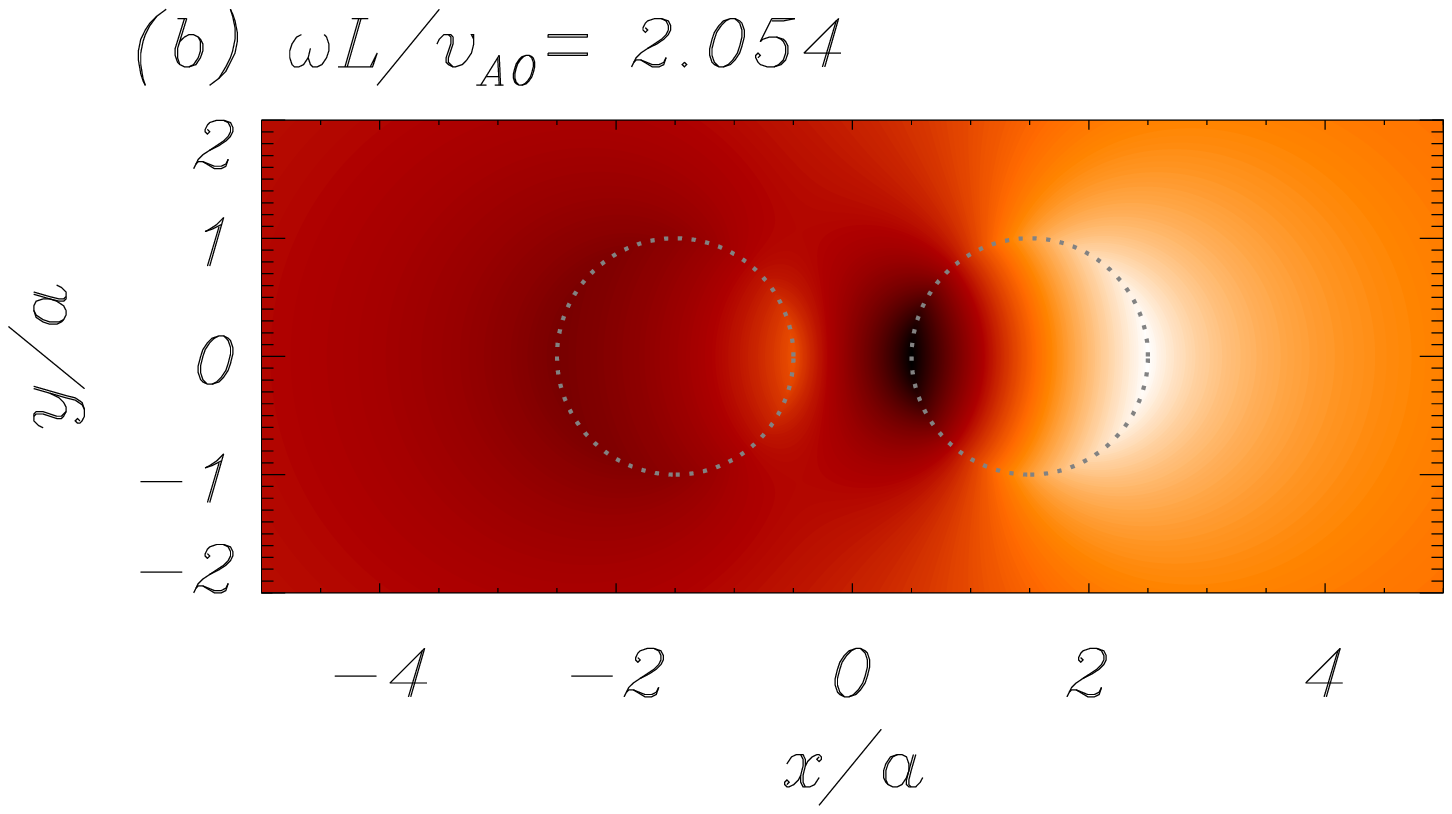}
\mbox{\hspace{-1.cm}\hspace{8.5cm}\hspace{8.5cm}\hspace{1.cm}}\vspace{-2.5cm}
\includegraphics[width=8.5cm]{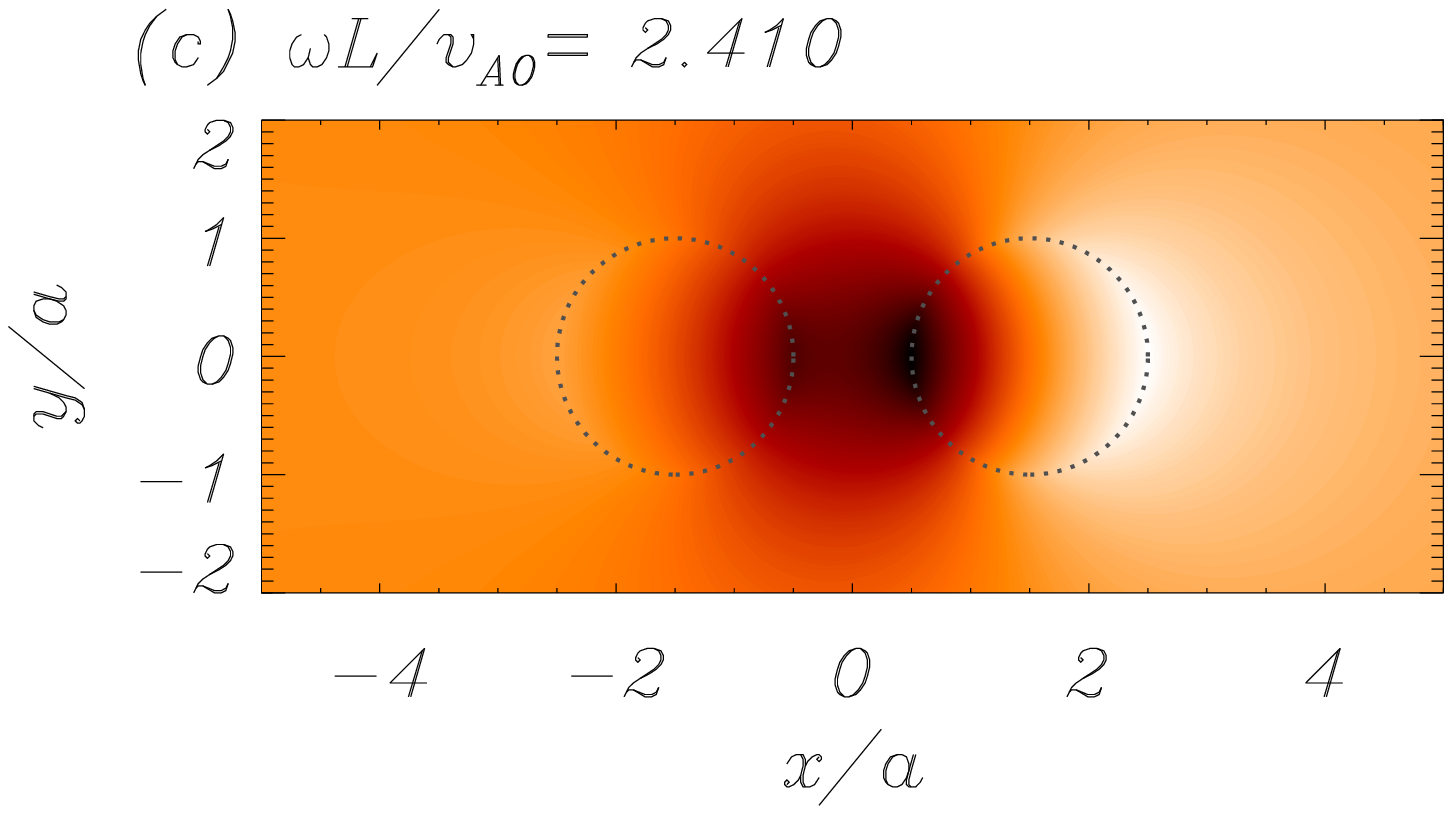}\hspace{-1.3cm}\includegraphics[width=8.5cm]{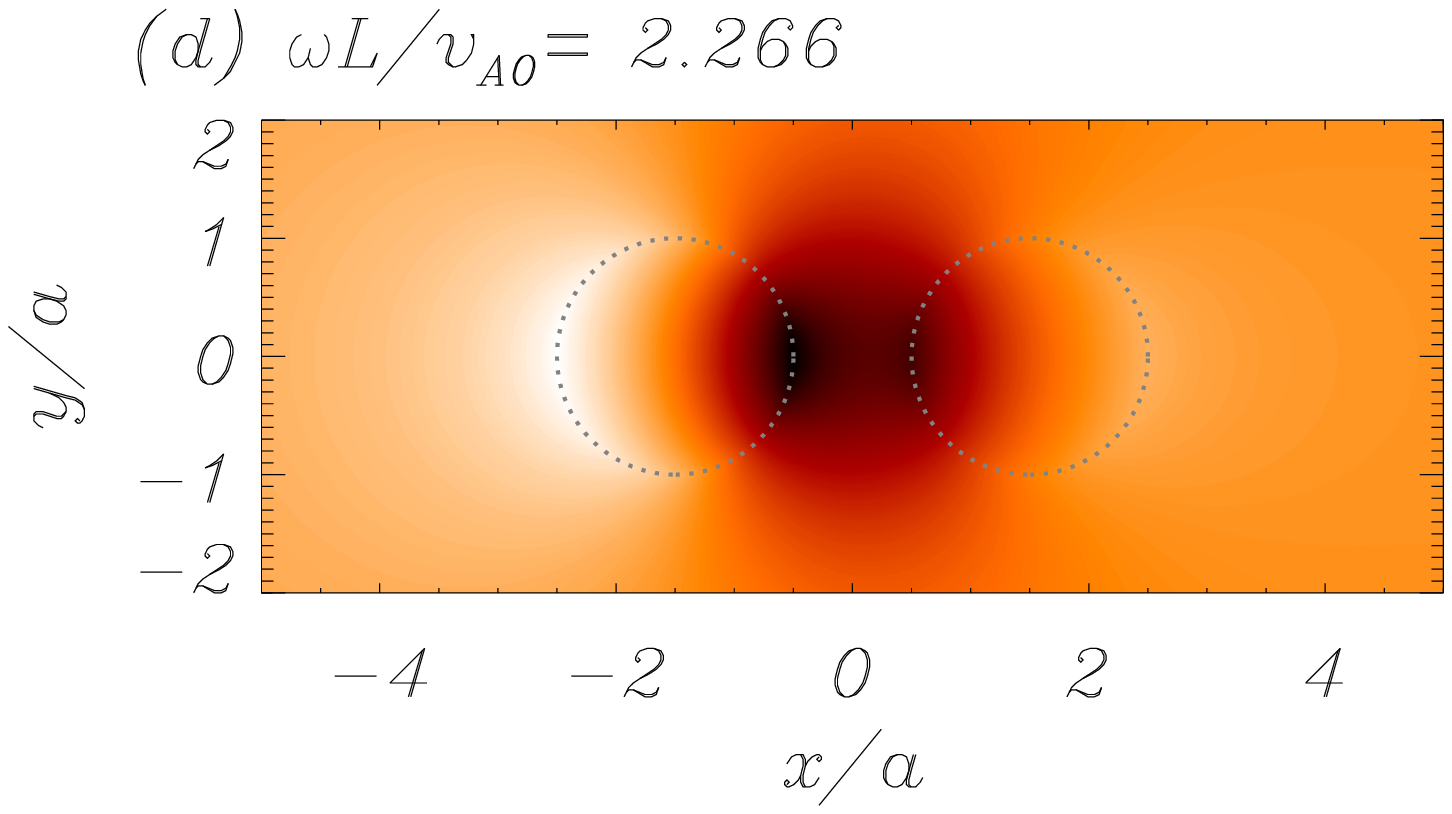}
\vspace{-2.5cm}
\center
\includegraphics[width=6.5cm]{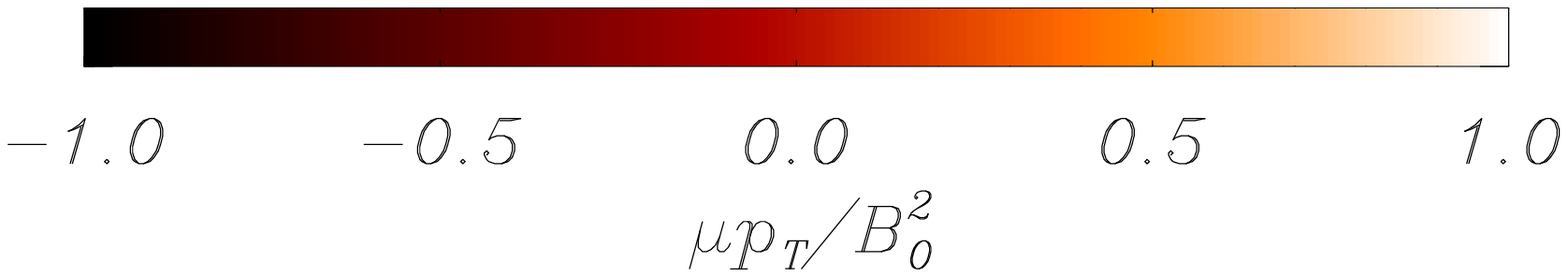}
\vspace{-1.5cm}
\caption{
Total pressure perturbation of the fast collective normal modes $P_x$ and $AP_x$
(plotted in the $xy$-plane) for a fixed density of the left loop
($\rho_\mathrm{1}=3\rho_\mathrm{0}$)  and different densities of the right loop
($\rho_\mathrm{2}$). The panels show the $P_x$ mode with {\bf (a)}
$\rho_\mathrm{2}=2.5\rho_\mathrm{0}$ and {\bf (b)}
$\rho_\mathrm{2}=3.5\rho_\mathrm{0}$; the $AP_x$ mode with {\bf (c)}
$\rho_\mathrm{2}=2.5\rho_\mathrm{0}$ and {\bf (d)}
$\rho_\mathrm{2}=3.5\rho_\mathrm{0}$. The frequencies of the modes are given on
top of the corresponding panels. The dotted lines show the boundaries of the
unperturbed tubes. Regions of the positive (negative) perturbed total pressure
represent density enhancements (decrements), so that in {\bf (a)} and {\bf (b)}
the loops move in phase in the $x$-direction, while in {\bf (c)} and {\bf (d)}
they move in antiphase in the $x$-direction.}
\label{2dif_loops}
\end{figure}

The total pressure perturbation of the $P_x$ and $AP_x$ modes is plotted in
Figure \ref{2dif_loops} for two cases in which the loop interaction is important
($\rho_\mathrm{2}=2.5 \rho_\mathrm{0}$ and $3.5 \rho_\mathrm{0}$). The behavior
of the other two modes, $AP_y$ and $P_y$, is analogous to that of the $P_x$ and
$AP_x$ modes and thus their spatial structure is not shown. In contrast to the
case $\rho_\mathrm{1}=\rho_\mathrm{2}$, in which the interaction is maximal and
thus the two loops oscillate with equal amplitudes \citep[see Fig.~$2$
of][]{luna2008} the solutions in Figure \ref{2dif_loops} display an imbalance in
the oscillatory amplitude of the two loops. The largest amplitude of the
pressure perturbation corresponds to the denser loop for the $P_x$ mode (see
Figs.~\ref{2dif_loops}a and \ref{2dif_loops}b), while it occurs in the rarer
loop for the $AP_x$ mode (see Figs.~\ref{2dif_loops}c and \ref{2dif_loops}d).

\begin{figure}[!ht]
\center
\resizebox{11.5cm}{!}{\includegraphics{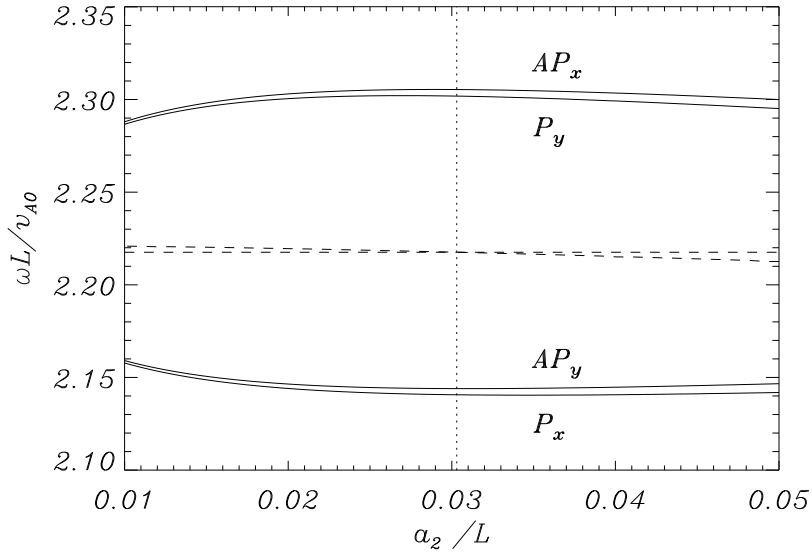}}
\caption{\small
Dimensionless frequency, $\omega L/v_\mathrm{A 0},$ of the four collective
kinklike modes $P_x$, $AP_y$, $P_y$, and $AP_x$ (solid lines), as a function of
the normalized right loop radius, $a_\mathrm{2}/L$. As in
Fig. \ref{ratio_densities}, the two individual kink frequencies are plotted
(dashed lines), where the horizontal dashed line corresponds to the left loop
and the one with a slight dependence on $a_\mathrm{2}$ corresponds to the right
loop.}
\label{ratio_radii}
\end{figure}

Secondly, we consider the same system of two loops but now the densities are
fixed to $\rho_\mathrm{1}=\rho_\mathrm{2}=3 \rho_\mathrm{0}$, the radius of the
left loop is $a_\mathrm{1}=0.03 L$, and the right loop radius, $a_\mathrm{2}$,
is allowed to vary. The distance between the tube centers is $3 a_\mathrm{M}$,
where $a_\mathrm{M}$ is the averaged radius defined as $a_\mathrm{M}=\left(
a_\mathrm{1}+a_\mathrm{2}\right) / 2$. With this condition the separation
measured in averaged radius units is constant. The frequencies of the four modes
$P_x$, $AP_y$, $P_y$, and $AP_x$ are plotted in Figure \ref{ratio_radii}. As in
Figure \ref{ratio_densities} the collective frequencies (solid lines) are
different from the individual kink frequencies (dashed lines), showing the
collective nature of the oscillations. The chosen range of radii are those
measured in TRACE observation of transverse oscillations \citep[see,
e.g.][]{aschwanden2003}. In Figure \ref{ratio_radii} we see that the collective
frequencies are more or less constant; moreover the amplitude of the oscillation
is more or less equal in each tube. Then, in the considered range of radii the
interaction between kink modes is strong and does not significantly depend on
the loop radii.

\section{Interaction between three loops}\label{3loops}

\subsection{Equal loop densities}\label{eq_loops}

\begin{figure}[!ht]
\center
\hspace{-1.5cm}
\mbox{\hspace{1.cm}\hspace{8.5cm}\hspace{8.5cm}\hspace{1.cm}}\vspace{-1.4cm}
\includegraphics[width=8.5cm]{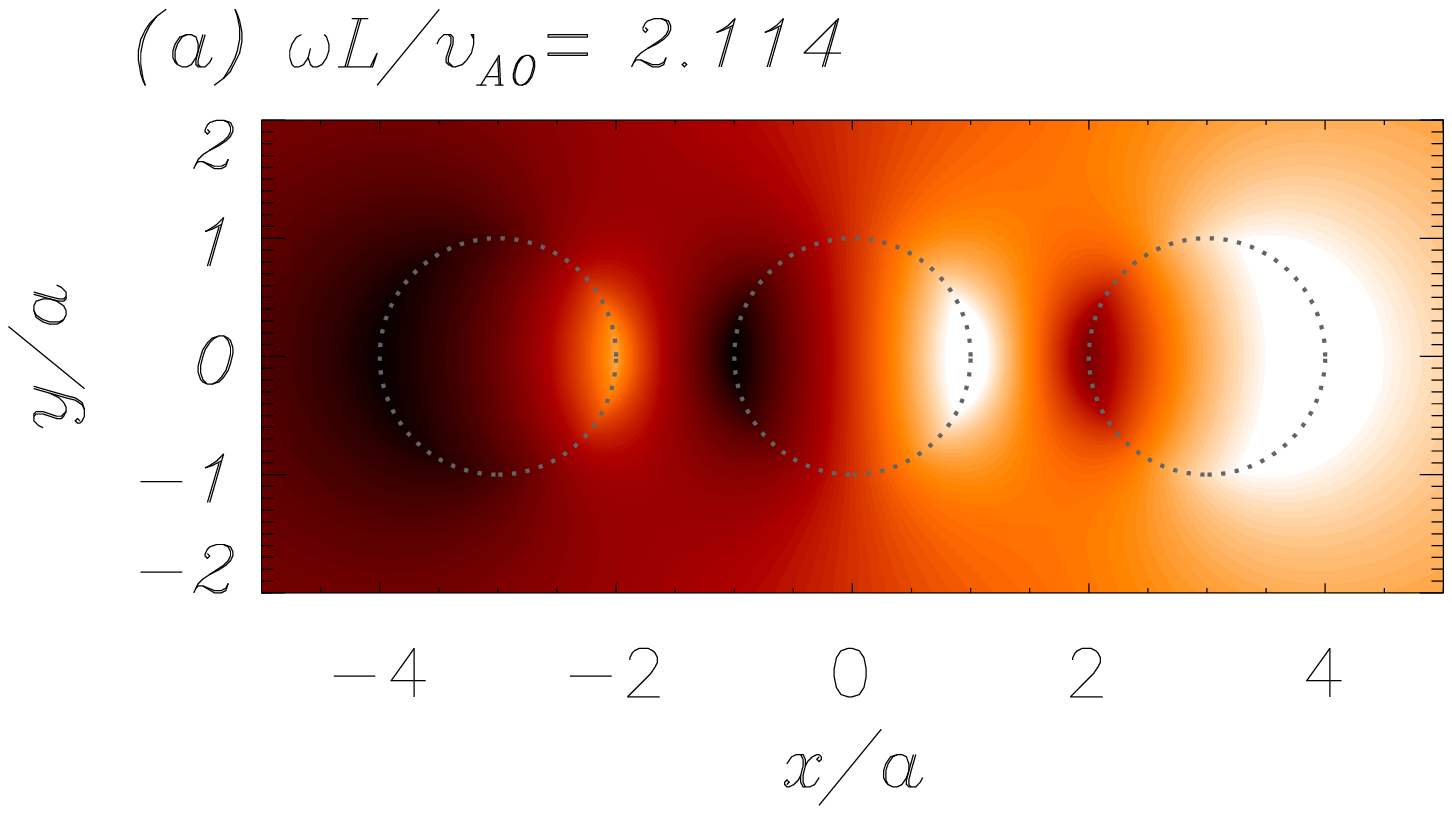}\hspace{-1cm}\includegraphics[width=8.5cm]{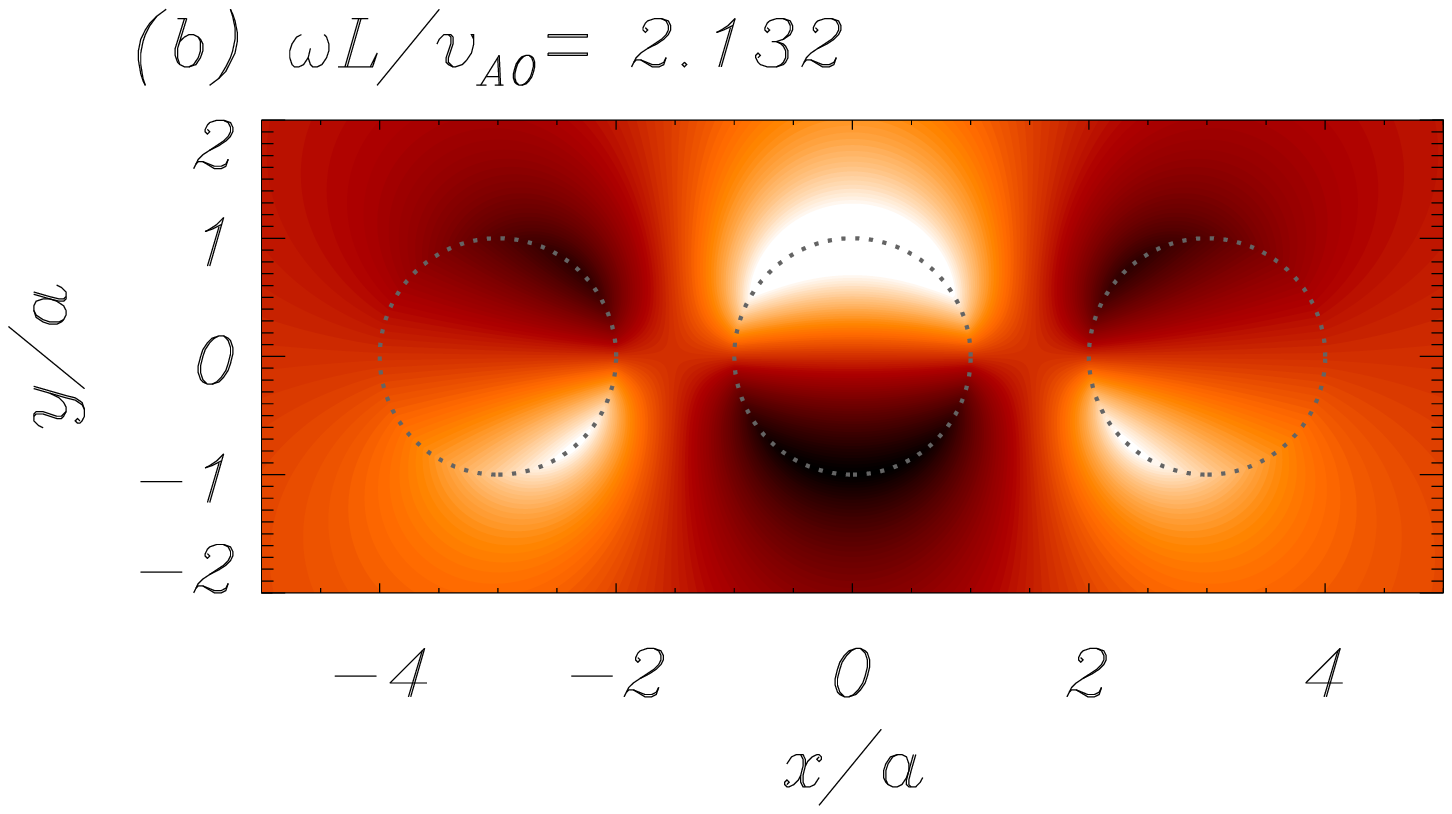}
\mbox{\hspace{1.cm}\hspace{4.5cm}\hspace{4.5cm}\hspace{1.cm}}\vspace{-2.4cm}
\includegraphics[width=8.5cm]{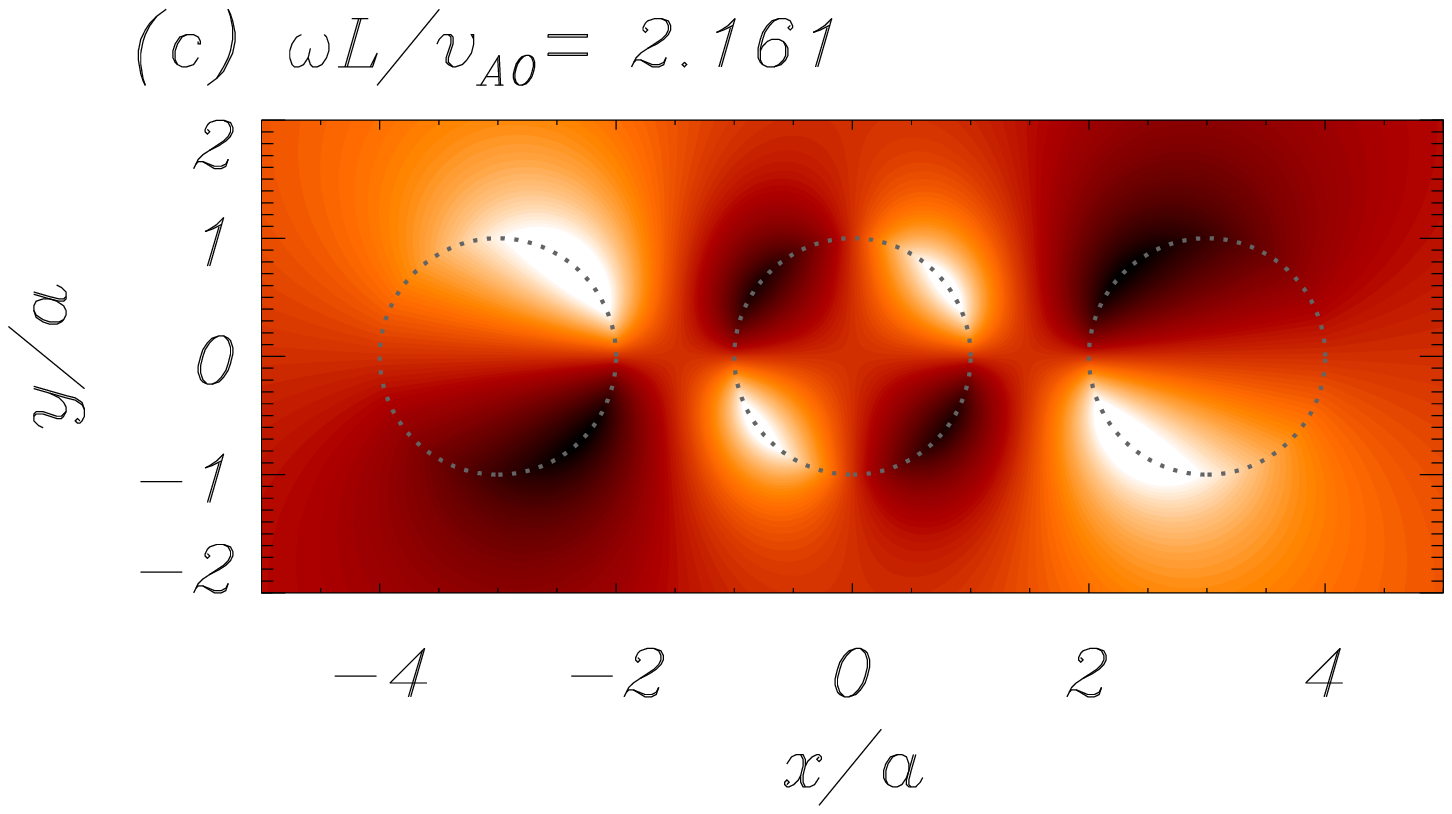}\hspace{-1cm}\includegraphics[width=8.5cm]{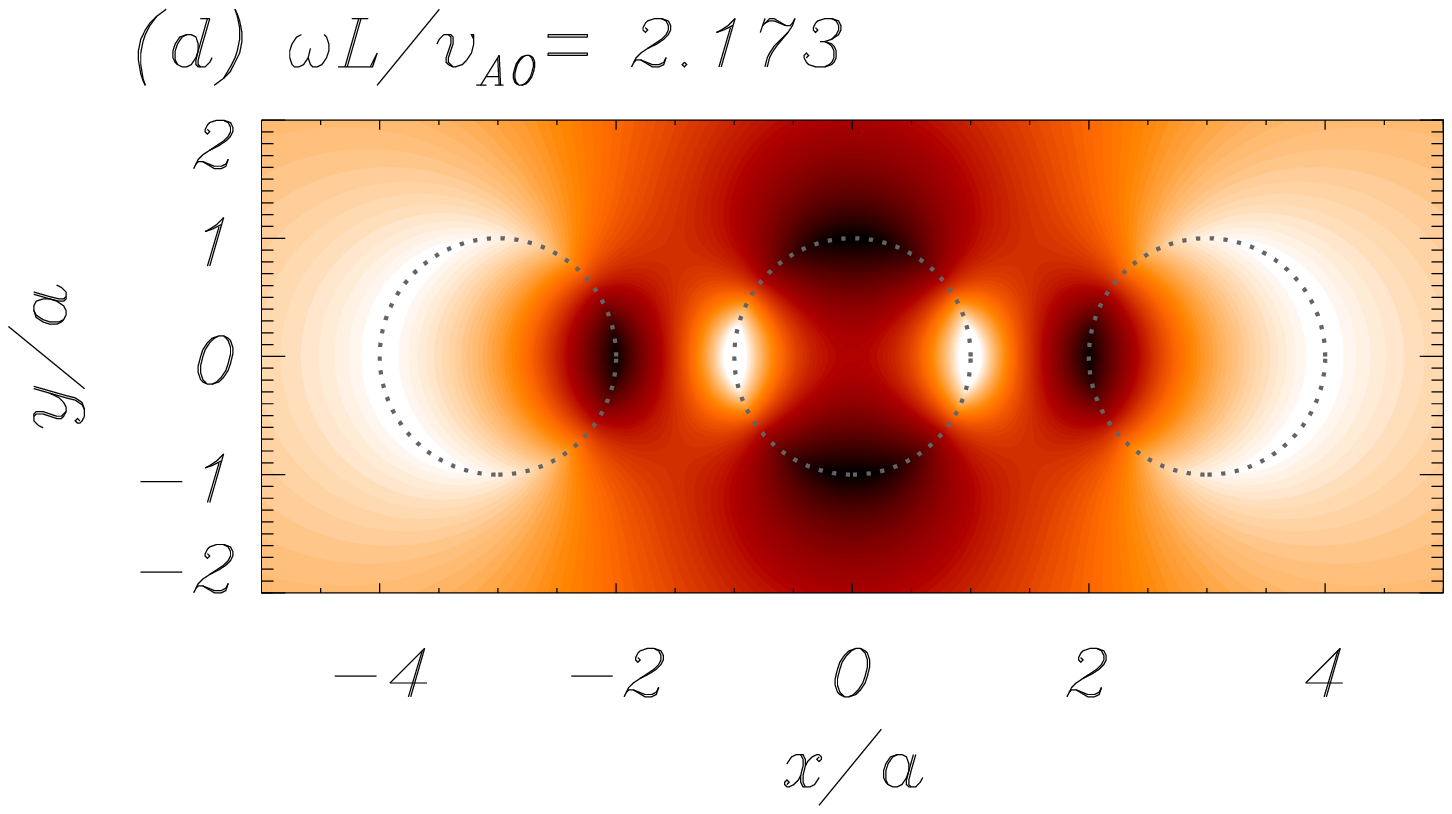}
\mbox{\hspace{1.cm}\hspace{4.5cm}\hspace{4.5cm}\hspace{1.cm}}\vspace{-2.4cm}
\includegraphics[width=8.5cm]{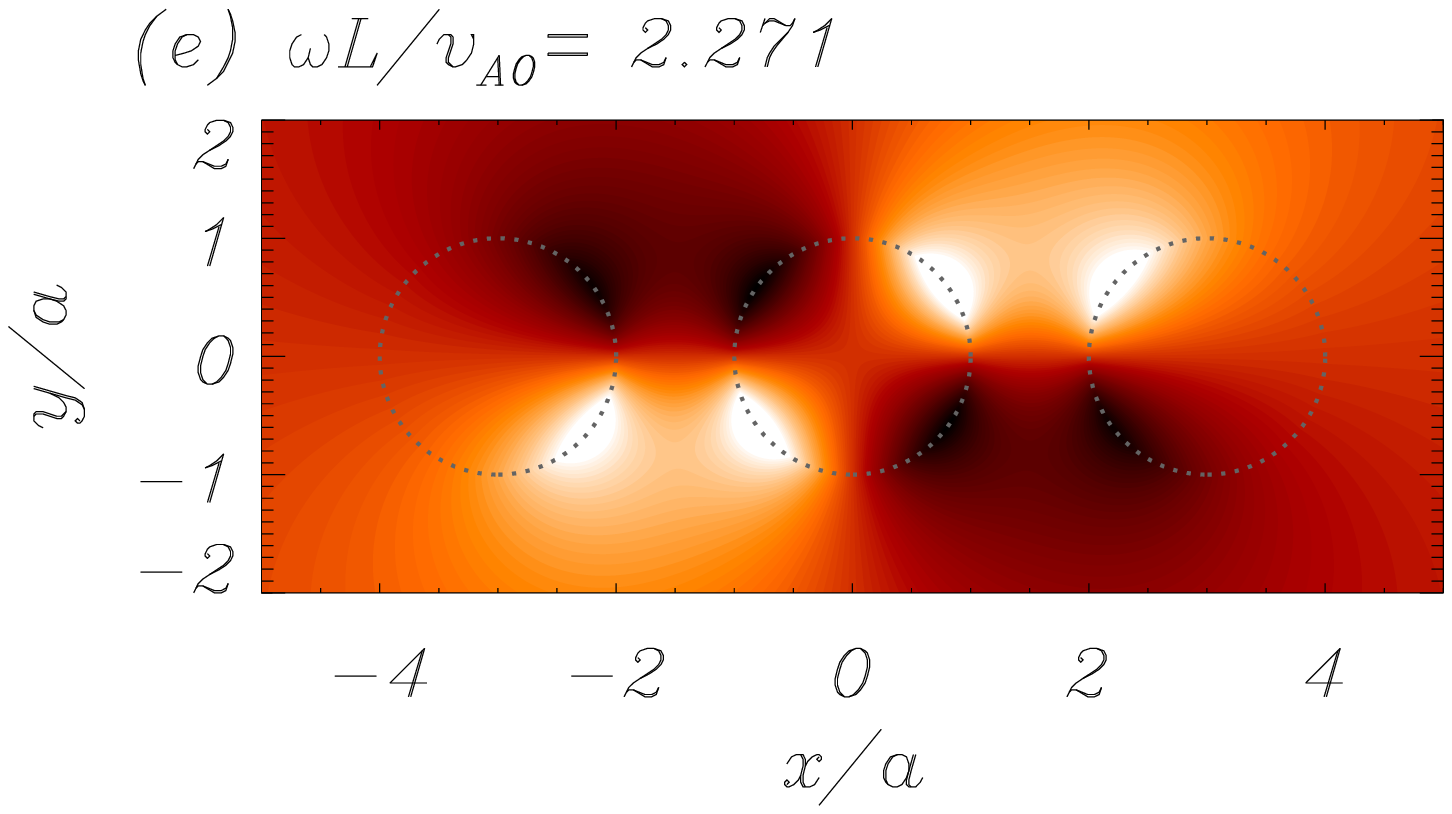}\hspace{-1cm}\includegraphics[width=8.5cm]{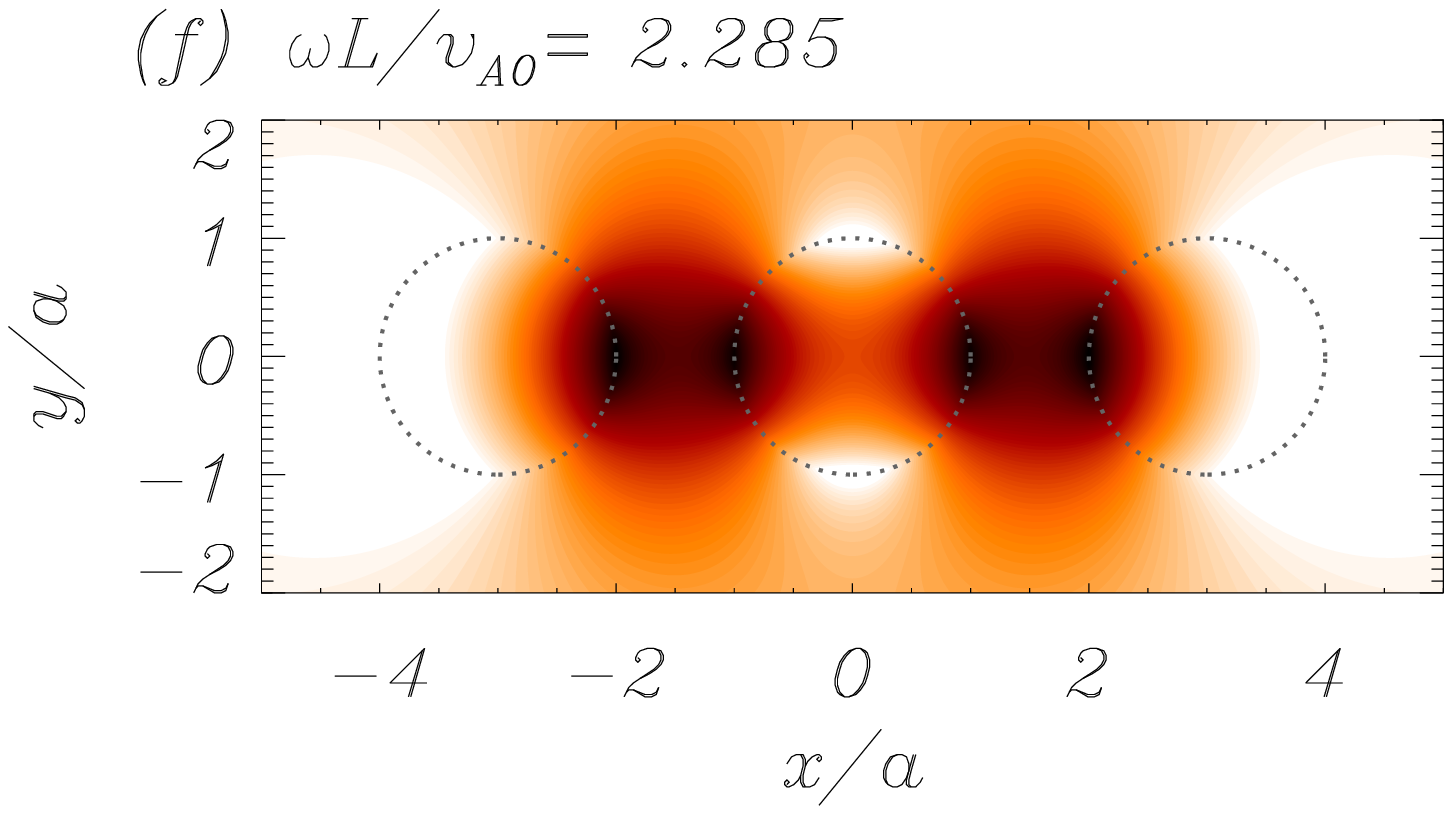}
\mbox{\hspace{1.cm}\hspace{4.5cm}\hspace{4.5cm}\hspace{1.cm}}\vspace{-2.4cm}
\includegraphics[width=8.5cm]{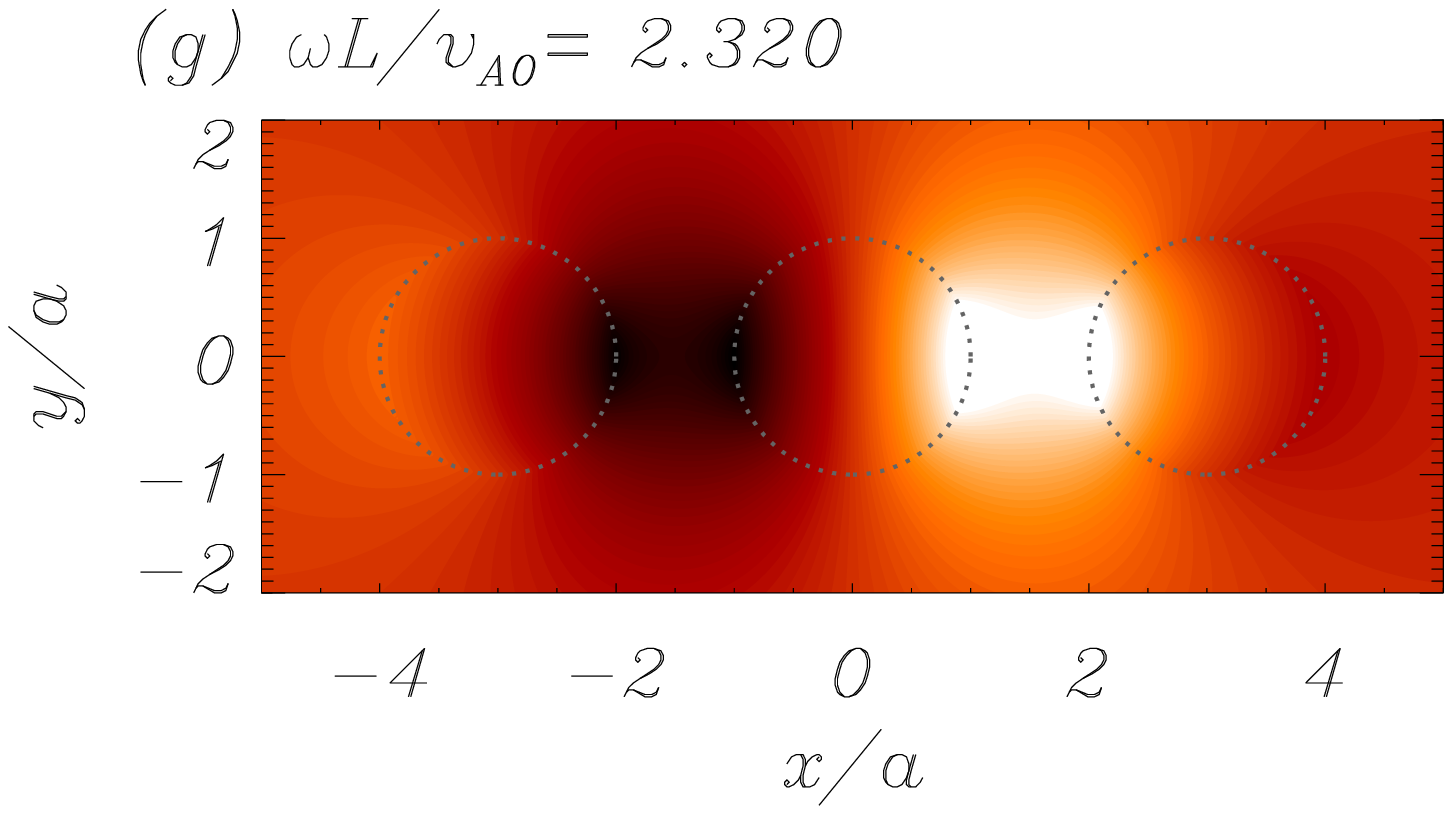}\hspace{-1cm}\includegraphics[width=8.5cm]{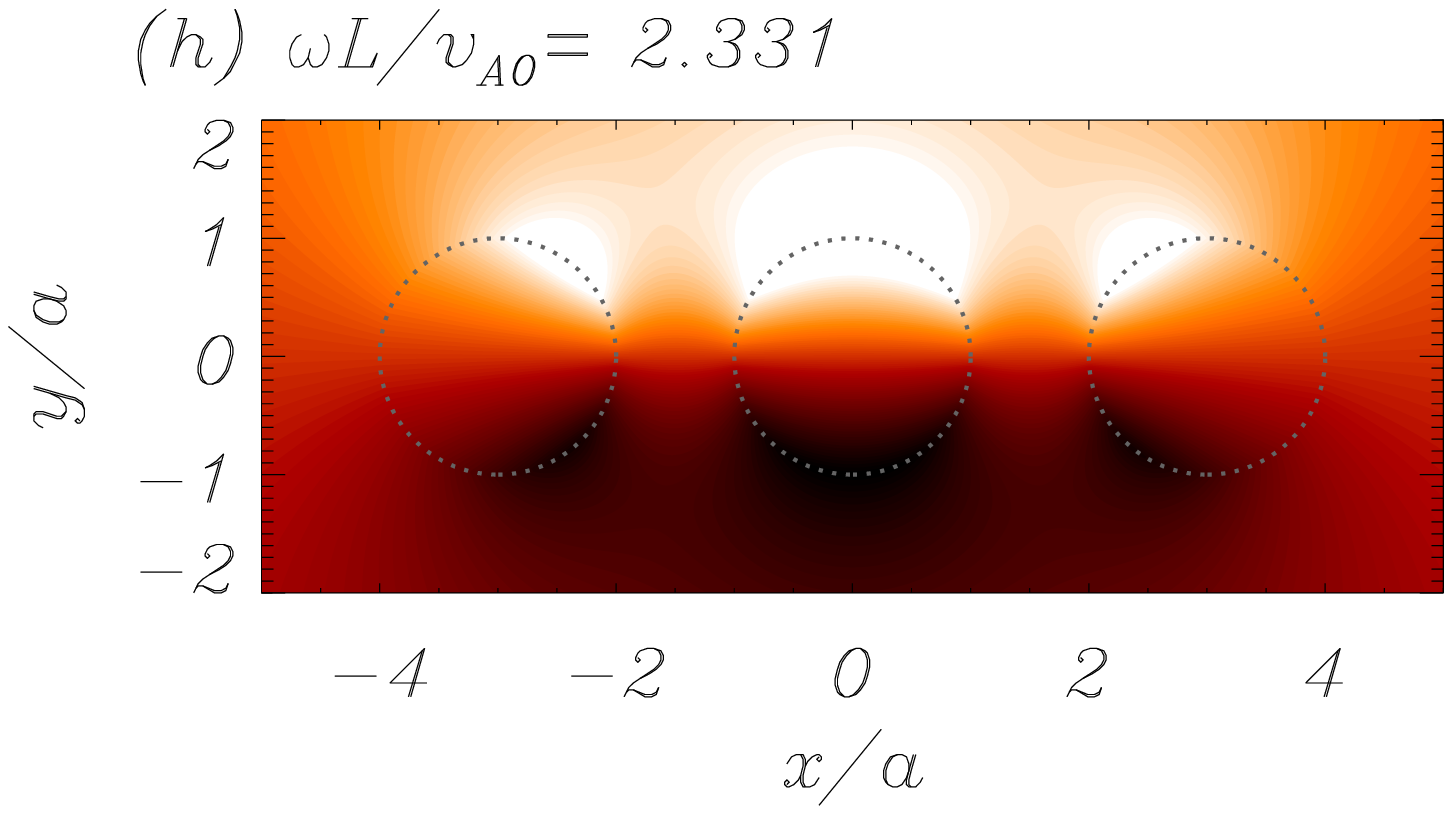}
\caption{
Total pressure perturbation of the eight kinklike collective normal modes of
three identical loops. The densities are fixed to
$\rho_\mathrm{1}=\rho_\mathrm{2}=\rho_\mathrm{3}=3 \rho_\mathrm{0}$, the radii
to $a_\mathrm{1}=a_\mathrm{2}=a_\mathrm{3}=a=0.03 L$, and the separation between
adjacent loops is $d=3 a$.}
\label{3loop_modes}
\end{figure}

We first study the situation in which the density and radii of the loops is the
same and find that there are eight kinklike normal modes, whose eigenfunctions
are plotted in Figure \ref{3loop_modes}, with the modes ordered by increasing
frequency. The lower frequency corresponds to a mode in which the three loops
move in phase in the $x$-direction (Fig.~\ref{3loop_modes}a), whereas in the
higher frequency mode (Fig.~\ref{3loop_modes}h) the three loops move in phase in
the $y$-direction. This behavior is different from that of the system of two
loops (see \S \ref{2loops}), in which the higher frequency mode corresponds to
the $A_x$ instead of the $S_y$ mode. The modes of Figures \ref{3loop_modes}a,
\ref{3loop_modes}b, \ref{3loop_modes}g, and \ref{3loop_modes}h are kinklike
while the other four modes of Figures \ref{3loop_modes}c, \ref{3loop_modes}d,
\ref{3loop_modes}e,  and \ref{3loop_modes}f combine kink and fluting
oscillations: the two left and right loops oscillate with a kinklike motion
whereas the central loop oscillates with a fluting motion. We also refer to
these modes as kinklike because at least one loop oscillates with a kinklike
behavior. In these modes the central loop contributes appreciably to the total
field (eq.\~[\ref{scattered_field_explicit}]) with the multipole $m=2$. Between
the frequencies of the modes plotted in Figures \ref{3loop_modes}d and
\ref{3loop_modes}e there are modes with the three loops oscillating with fluting
motions and even with more complex structure associated to $m>2$ solutions. Then
we call these modes flutinglike. They are not further analysed because they do
not produce transverse displacements of the loops.

\begin{figure}[!ht]
\center
\resizebox{11.5cm}{!}{\includegraphics{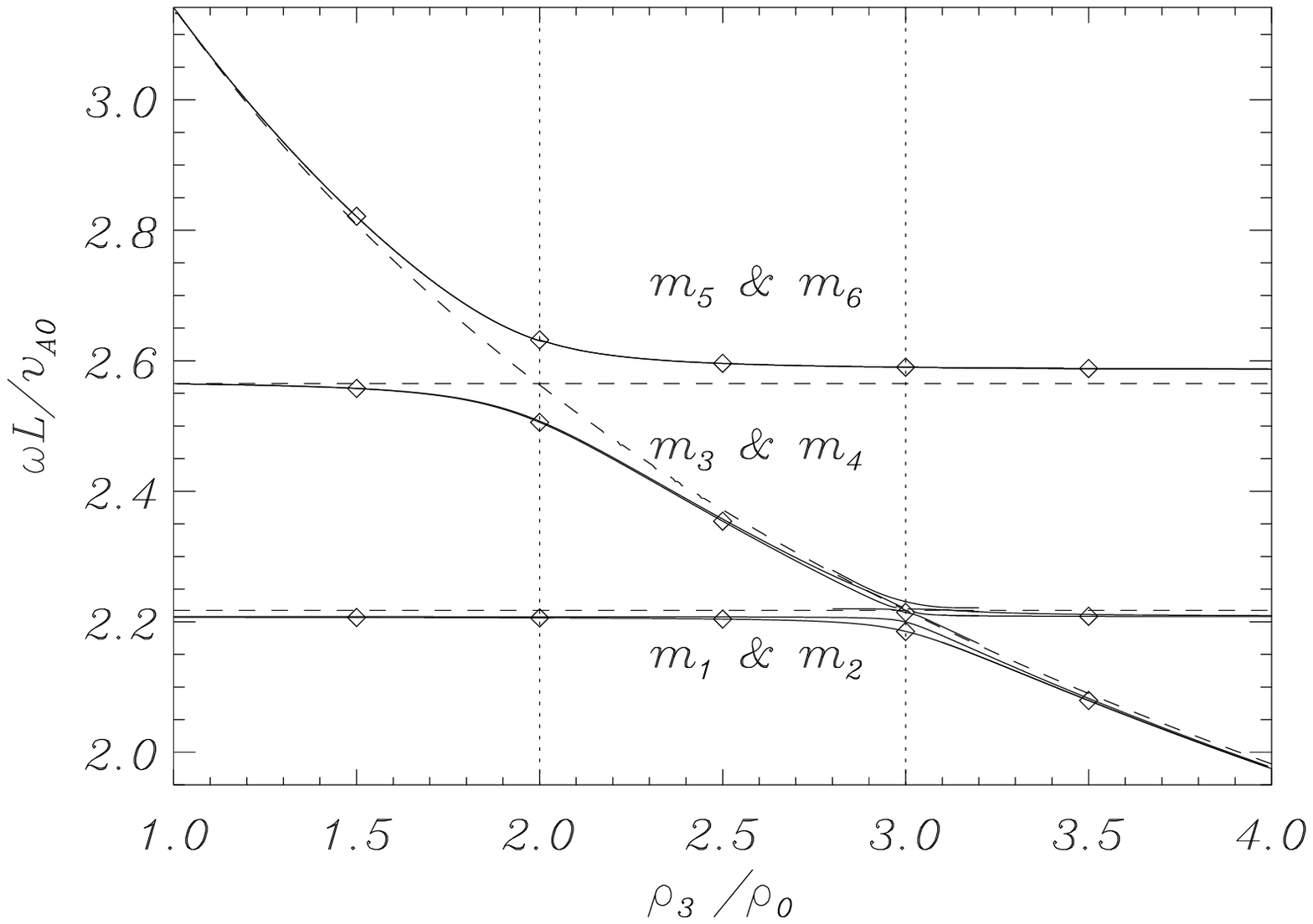}}
\caption{
\small Same as Fig. \ref{ratio_densities} for the collective frequencies of
three aligned loops plotted as a function of the density of loop $3$. Solid
lines correspond to the frequencies of the six collective kinklike modes. Dashed
lines correspond to the individual kink frequencies of the loops. The horizontal
bottom and upper dashed lines corresponds to the kink frequencies of loops
$\mathrm{1}$ and $\mathrm{2}$ respectively. The other dashed curve corresponds
to the kink frequency of the third loop, with variable density
$\rho_\mathrm{3}$. Diamonds mark the frequencies of the modes represented in
Figs. \ref{3nonidloops_densities_noncoupled} and
\ref{3nonidloops_densities_coupled}.}
\label{3loops_densities}
\end{figure}

\subsection{Different loop densities}\label{dif_loops}

\begin{figure}[!ht]
\center
\mbox{\hspace{1.cm}\hspace{7.cm}\hspace{7.cm}\hspace{7.cm}\hspace{3.cm}}\vspace{-1cm}
\hspace{-1.cm}\includegraphics[width=6cm]{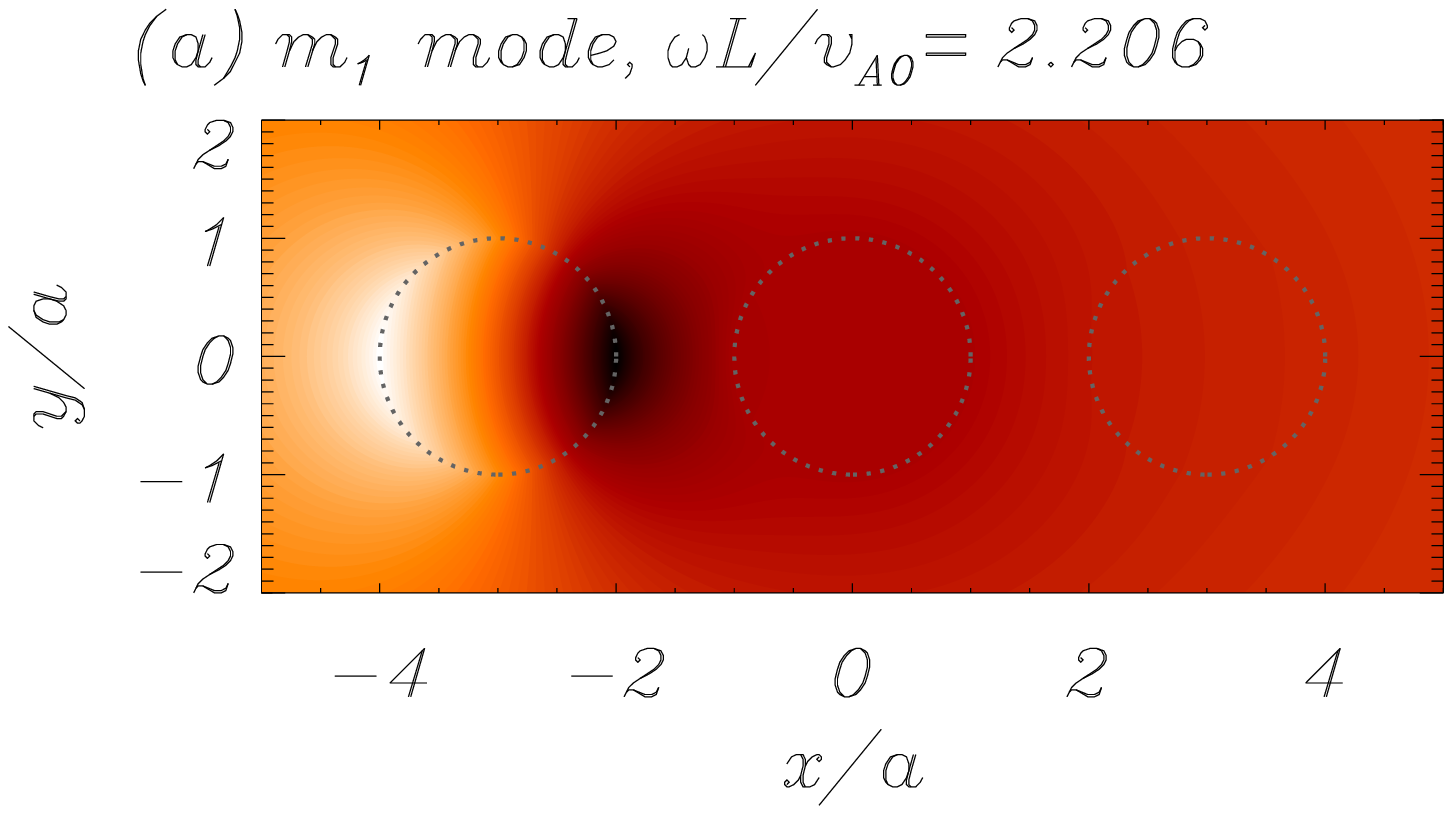}\hspace{-0.8cm}\includegraphics[width=6cm]{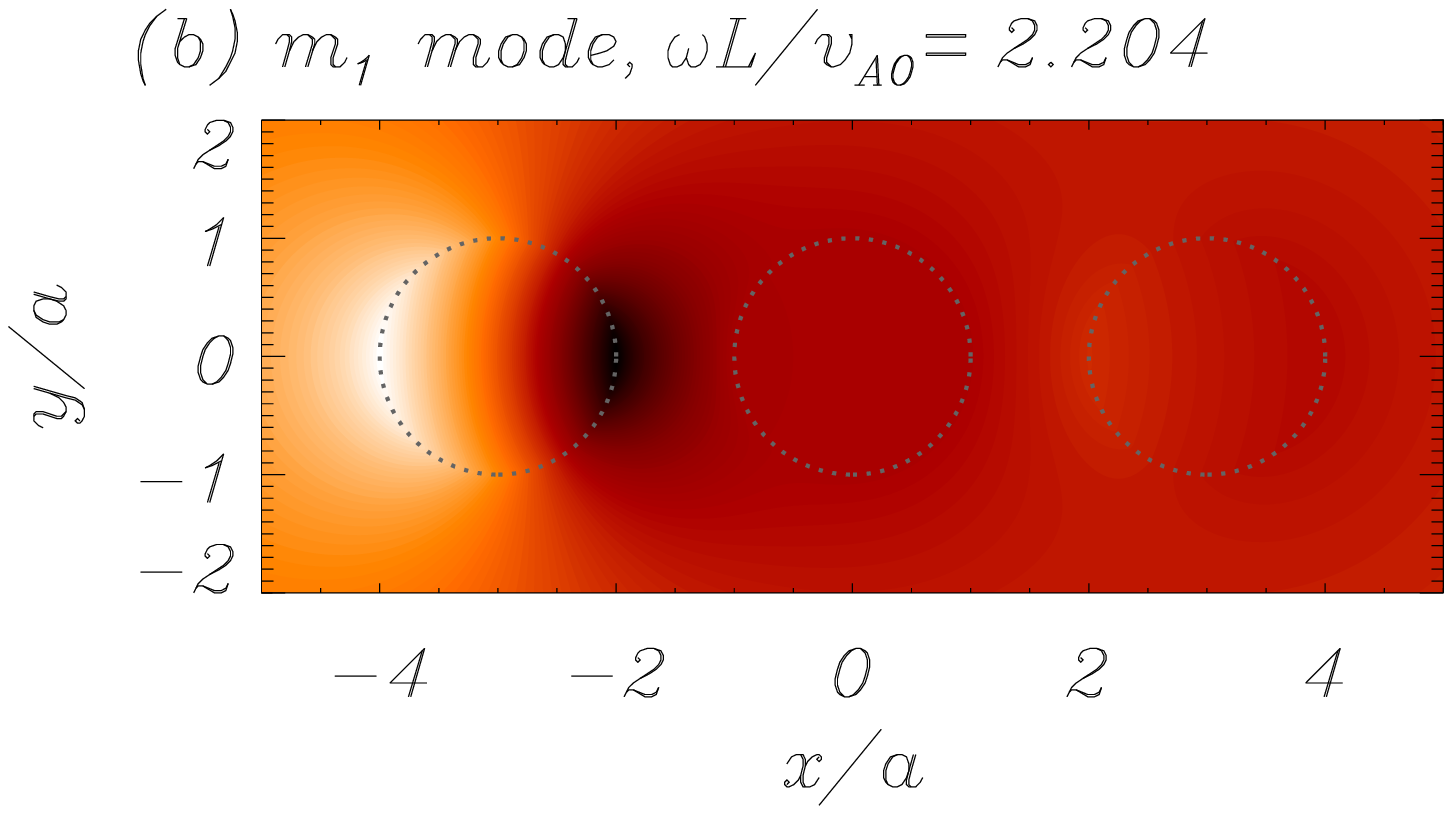}\hspace{-0.8cm}\includegraphics[width=6cm]{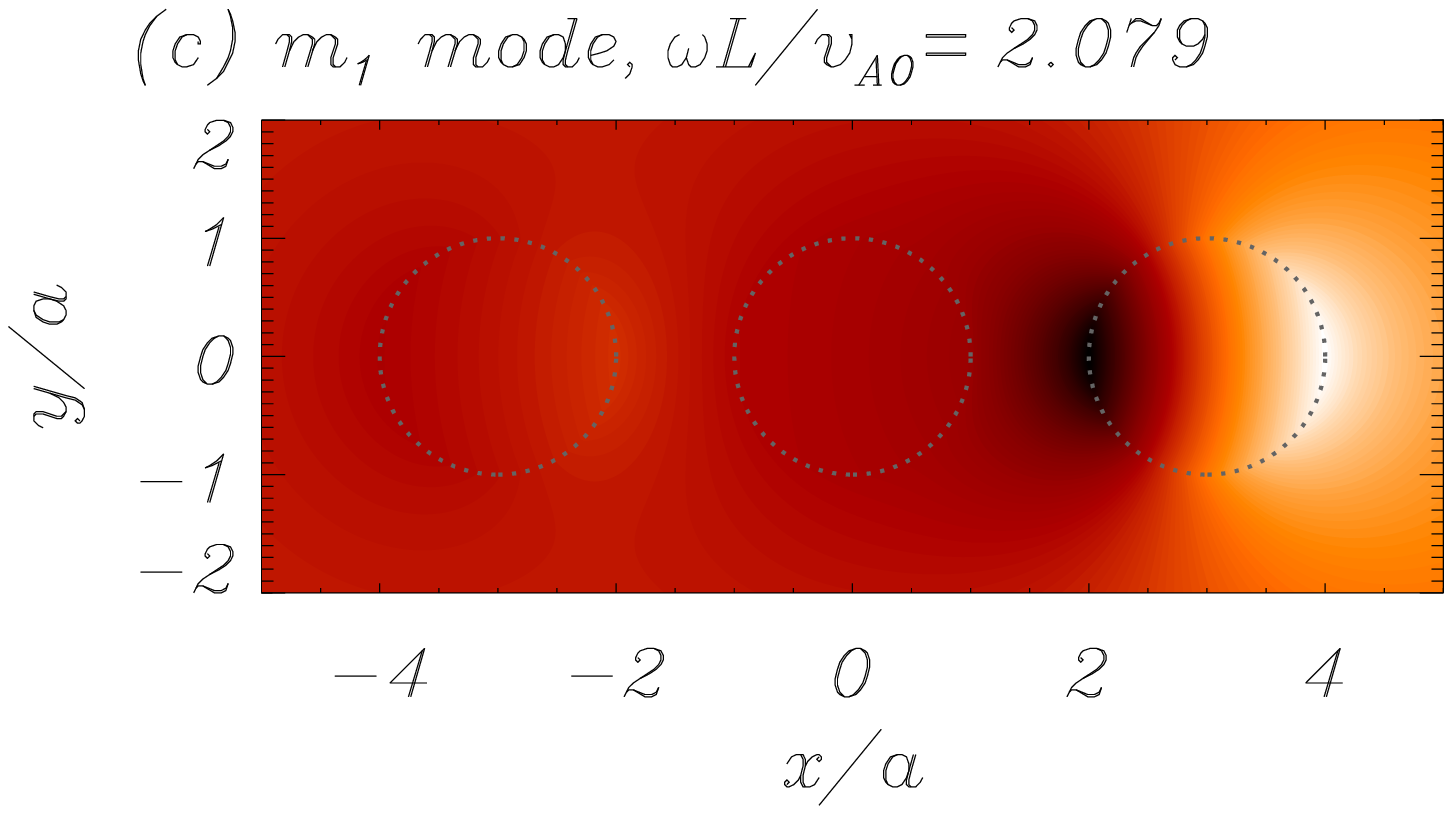}\vspace{-1cm}
\mbox{\hspace{1.cm}\hspace{7.cm}\hspace{7.cm}\hspace{7.cm}\hspace{3.cm}}\vspace{-1cm}
\hspace{-1.cm}\includegraphics[width=6cm]{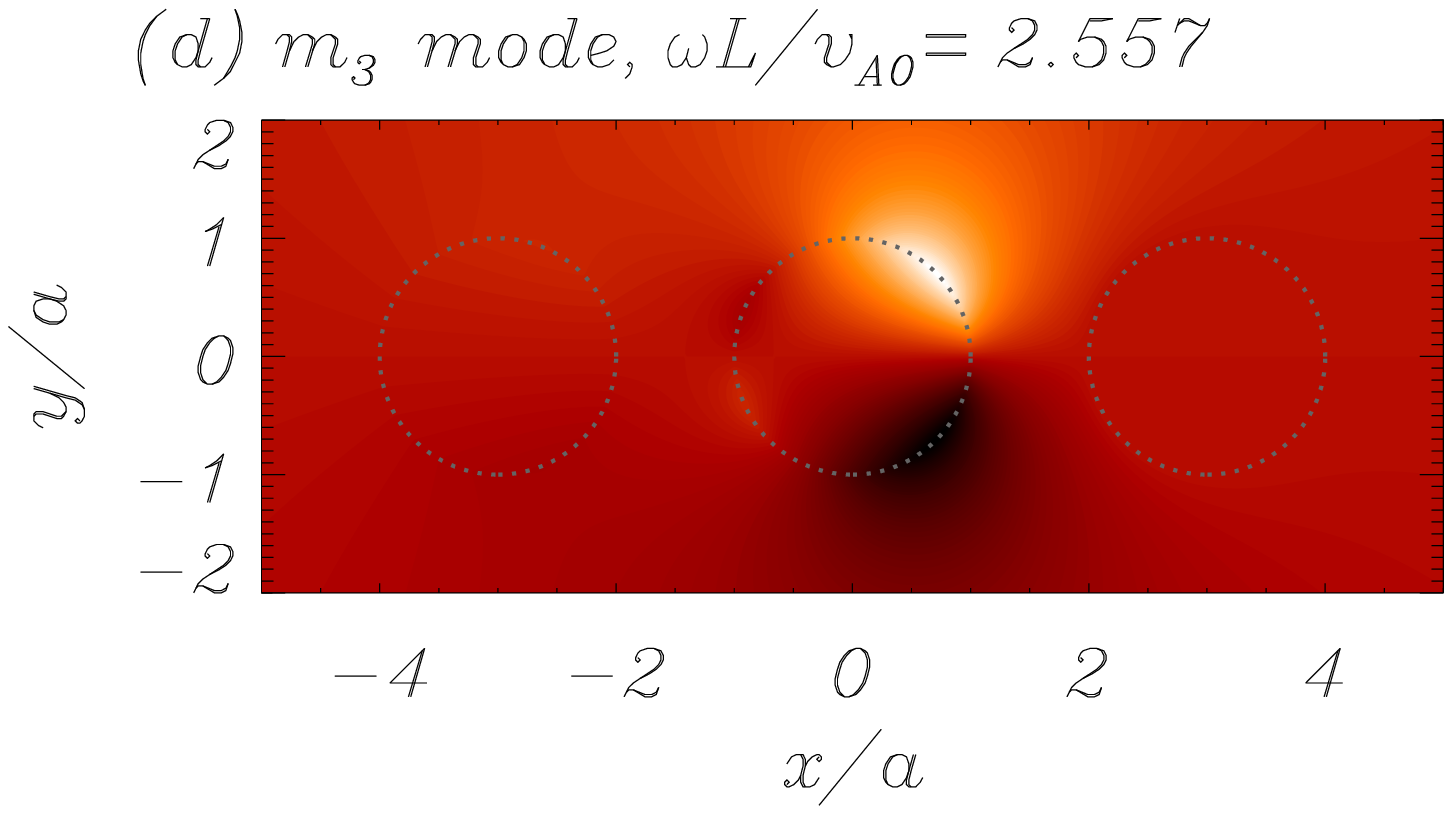}\hspace{-0.8cm}\includegraphics[width=6cm]{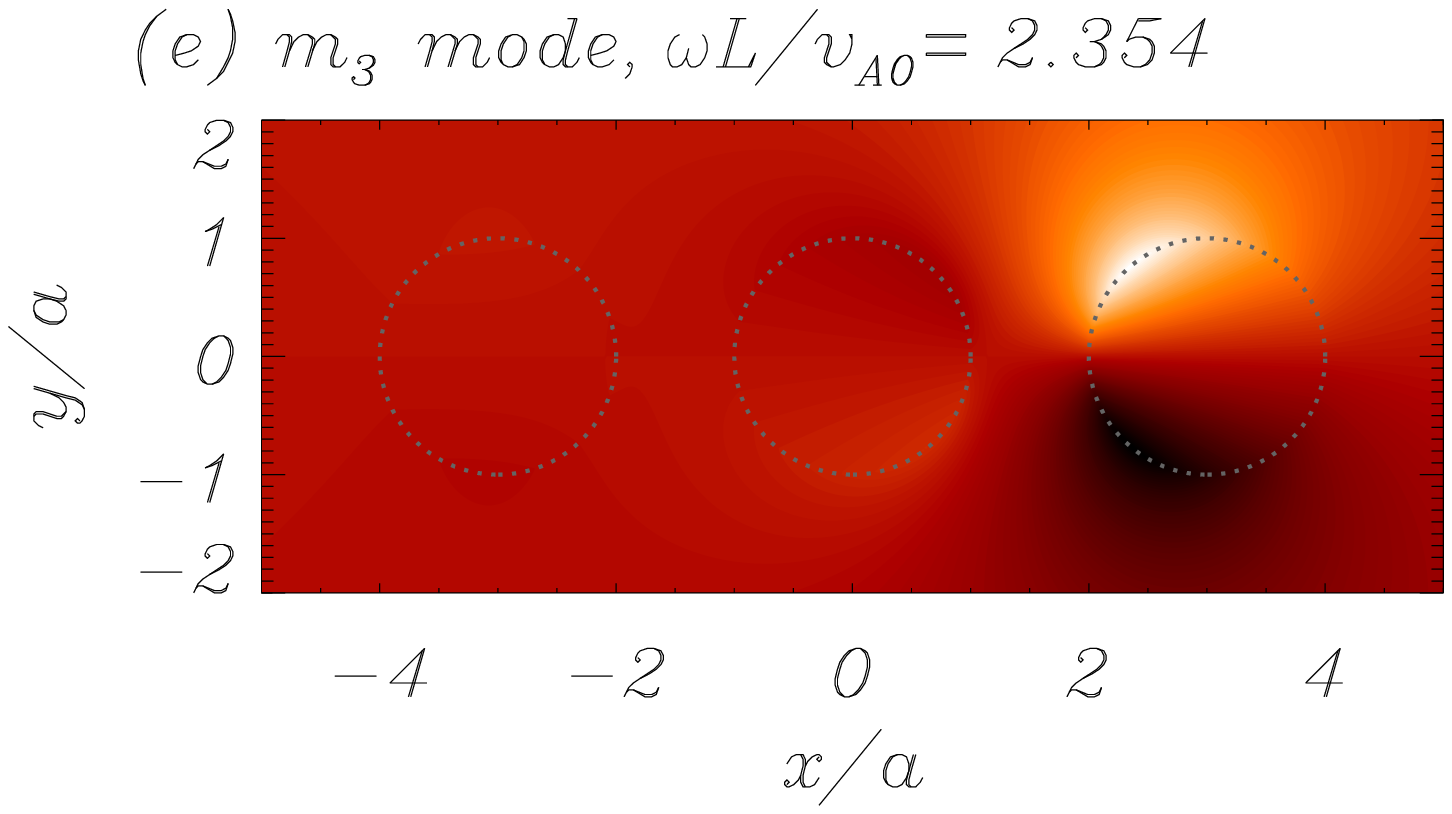}\hspace{-0.8cm}\includegraphics[width=6cm]{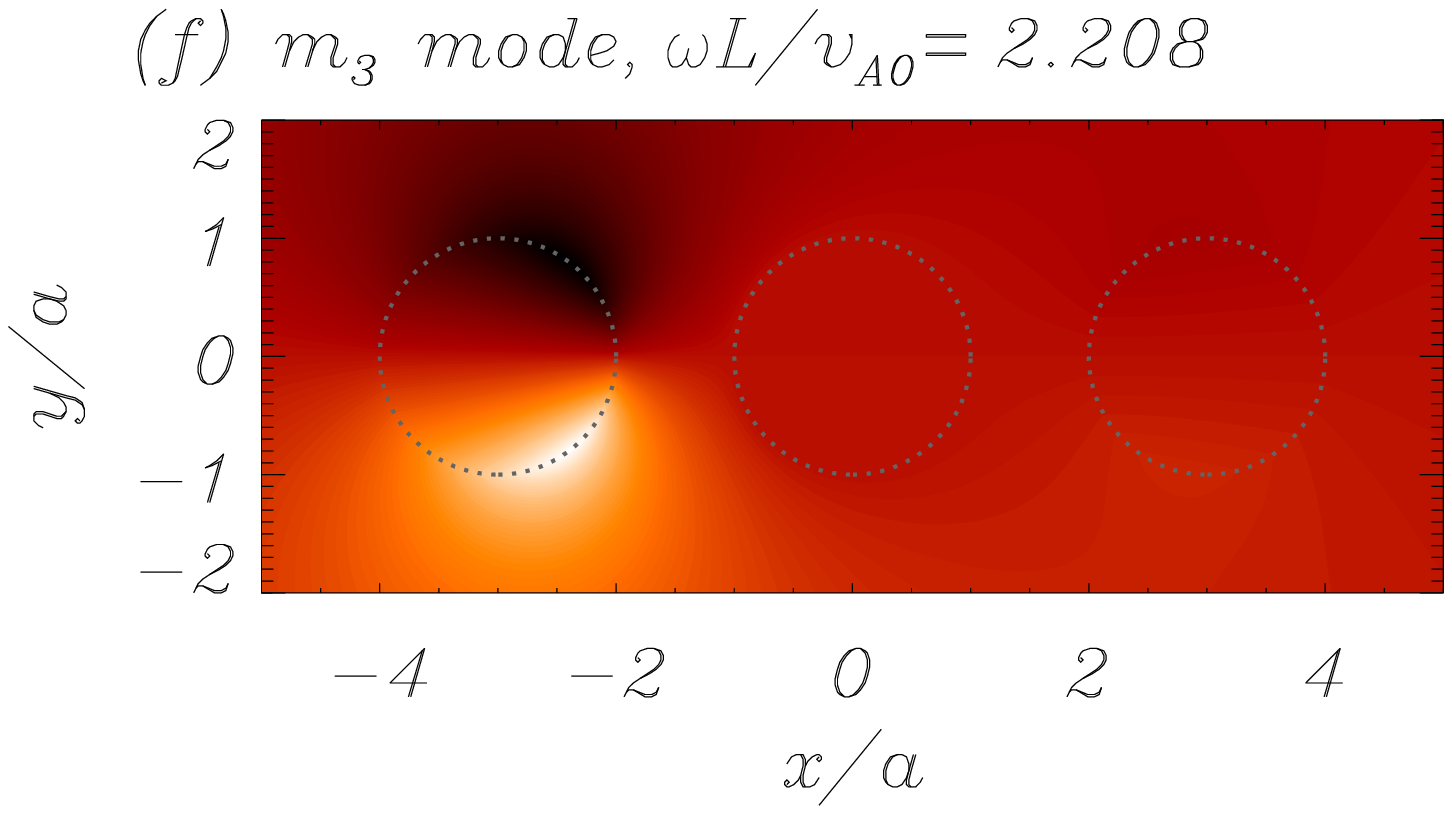}\vspace{-1cm}
\mbox{\hspace{1.cm}\hspace{7.cm}\hspace{7.cm}\hspace{7.cm}\hspace{3.cm}}\vspace{-1cm}
\hspace{-1.cm}\includegraphics[width=6cm]{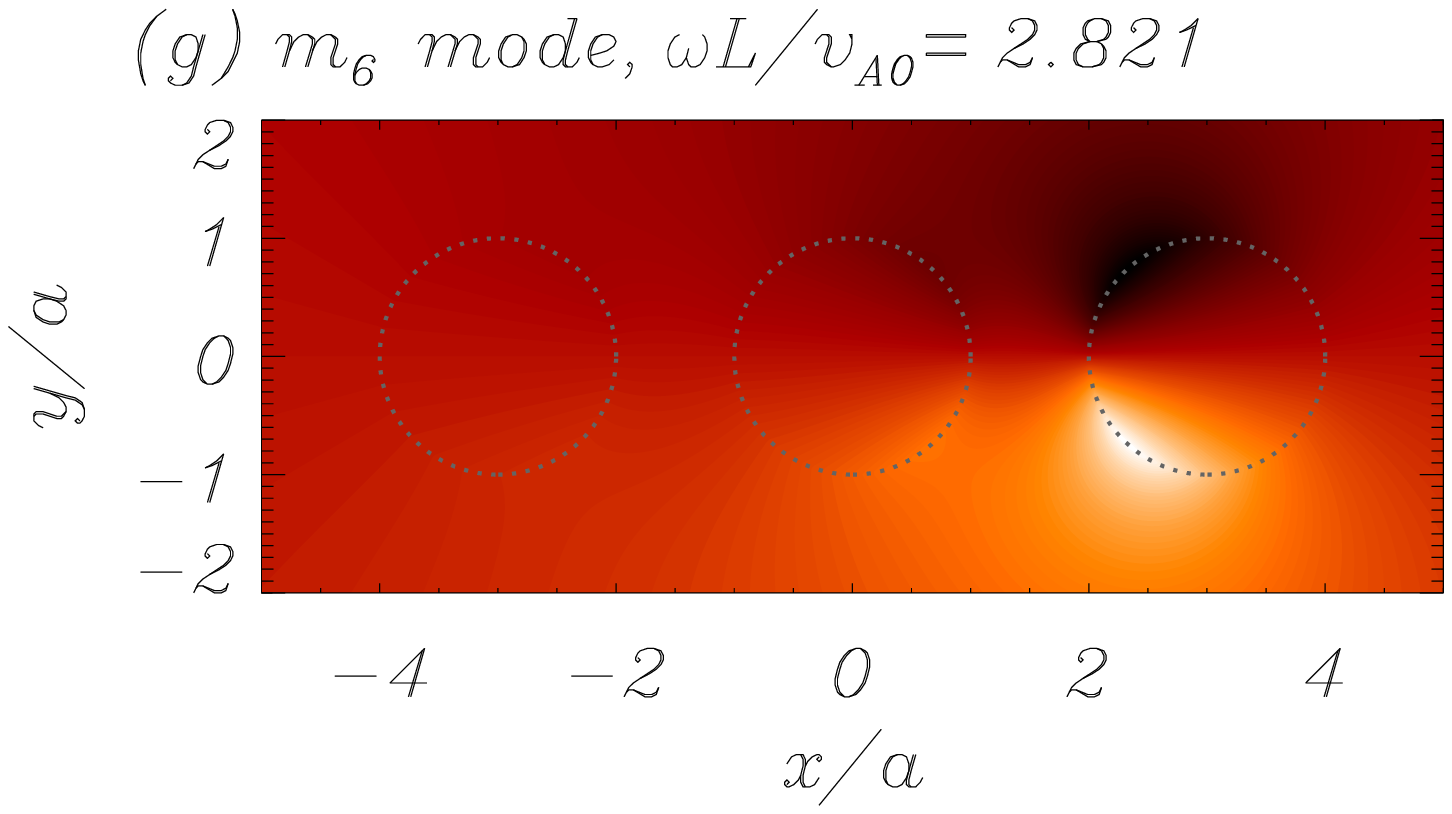}\hspace{-0.8cm}\includegraphics[width=6cm]{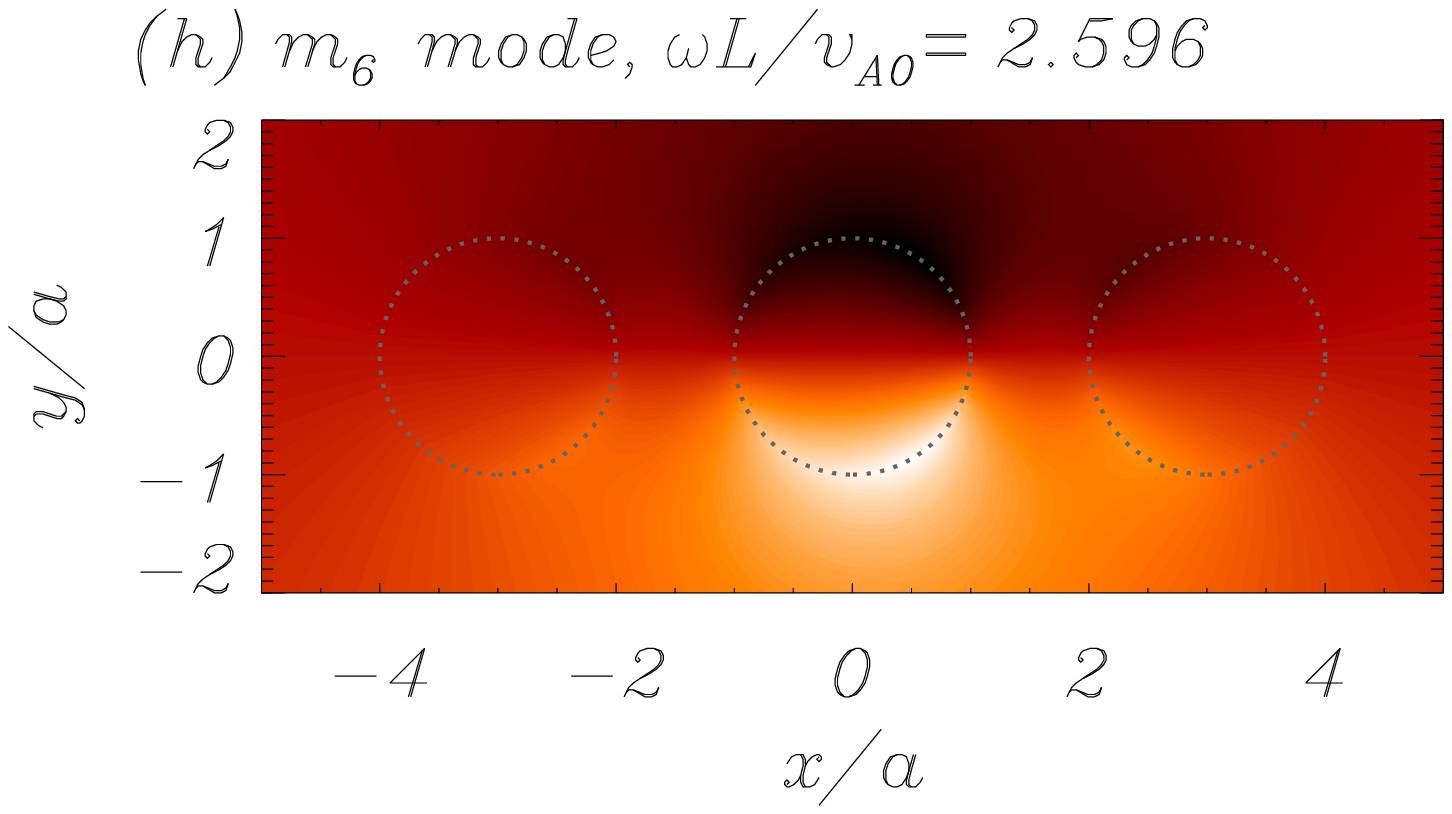}\hspace{-0.8cm}\includegraphics[width=6cm]{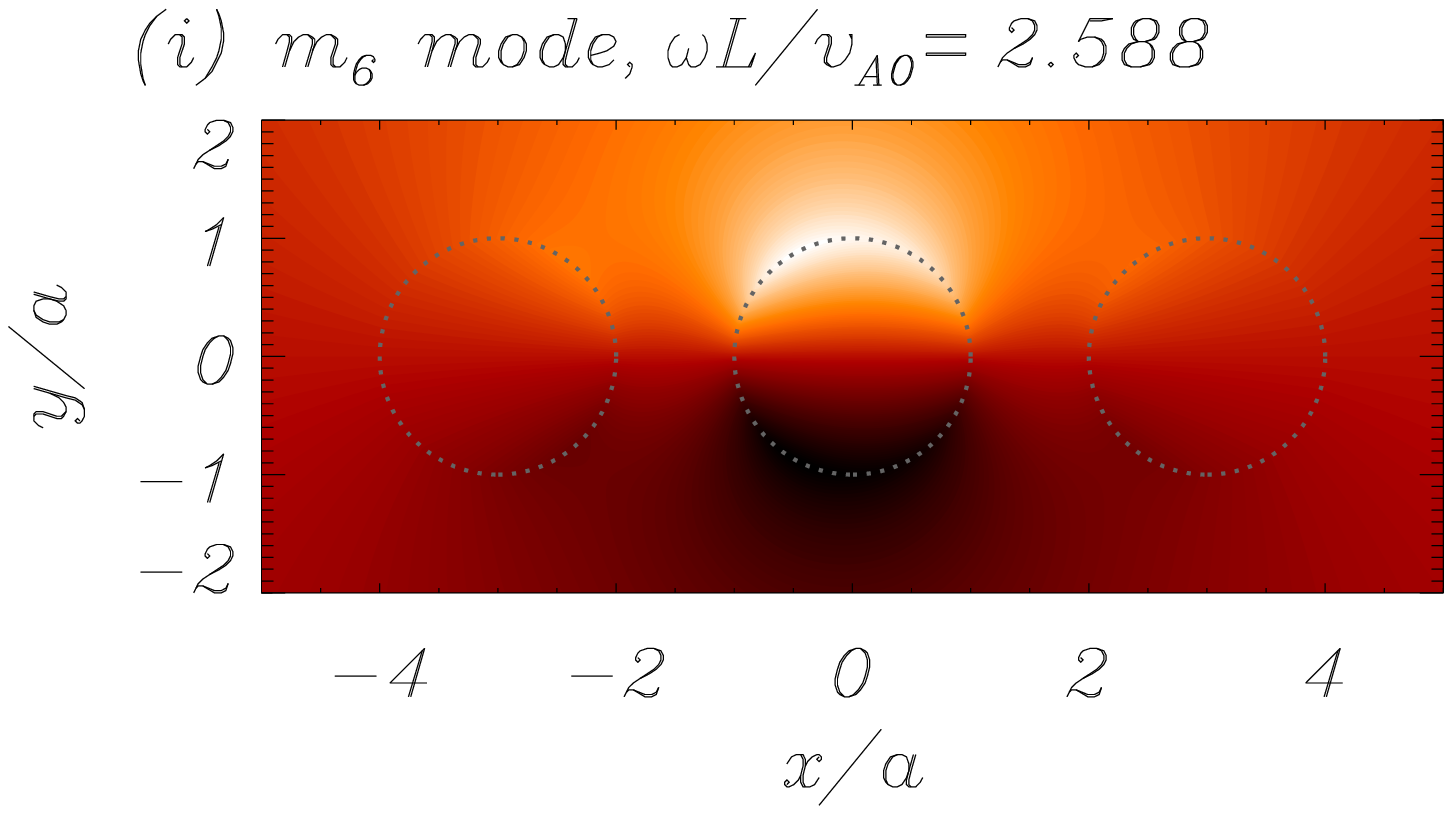}
\caption{
Same as Fig. \ref{3loop_modes} for three values of $\rho_\mathrm{3}$ far from
the two coupling regions of Fig. \ref{3loops_densities}. The densities of
loops $1$ and $2$ are fixed to $\rho_\mathrm{1}=3 \rho_\mathrm{0}$ and
$\rho_\mathrm{2}=2 \rho_\mathrm{0}$ and $\rho_\mathrm{3}$ is changed. The loop
radii are fixed to $a_\mathrm{1}=a_\mathrm{2}=a_\mathrm{3}=0.03 L$. In the top,
central, and bottom rows of panels the $m_\mathrm{1}$, $m_\mathrm{3}$, and
$m_\mathrm{6}$ modes are plotted for {\bf (a)}, {\bf (d)}, and  {\bf (g)}
$\rho_\mathrm{3}=1.5 \rho_\mathrm{0}$; {\bf (b)}, {\bf (e)}, and  {\bf (h)}
$\rho_\mathrm{3}=2.5 \rho_\mathrm{0}$; {\bf (c)}, {\bf (f)}, and  {\bf (i)}
$\rho_\mathrm{3}=3.5 \rho_\mathrm{0}$.}
\label{3nonidloops_densities_noncoupled}
\end{figure}

\begin{figure}[!ht]
\center
\mbox{\hspace{1.cm}\hspace{8.5cm}\hspace{8.5cm}\hspace{1.cm}}\vspace{-1.4cm}
\includegraphics[width=8.5cm]{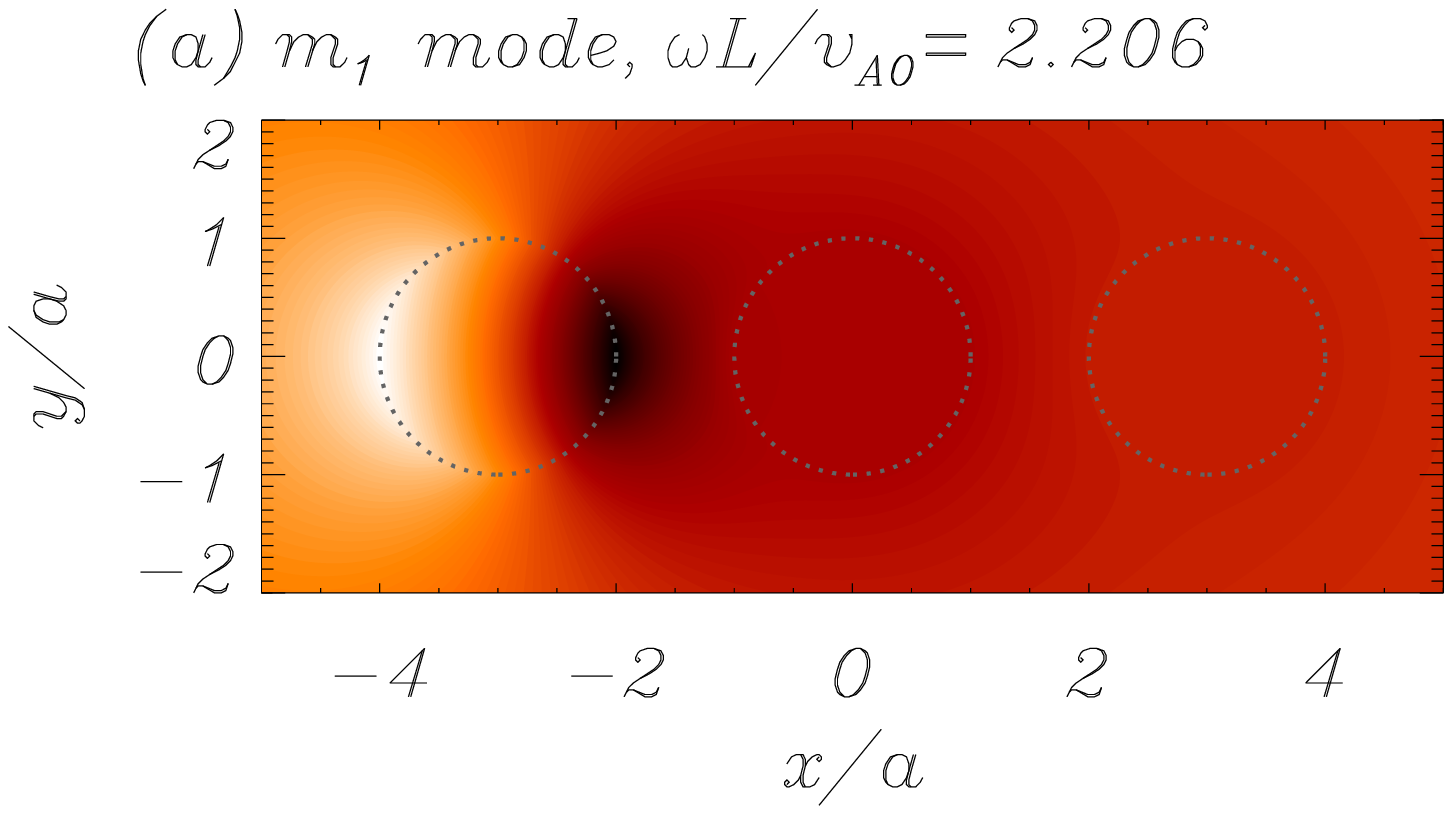}\hspace{-1cm}\includegraphics[width=8.5cm]{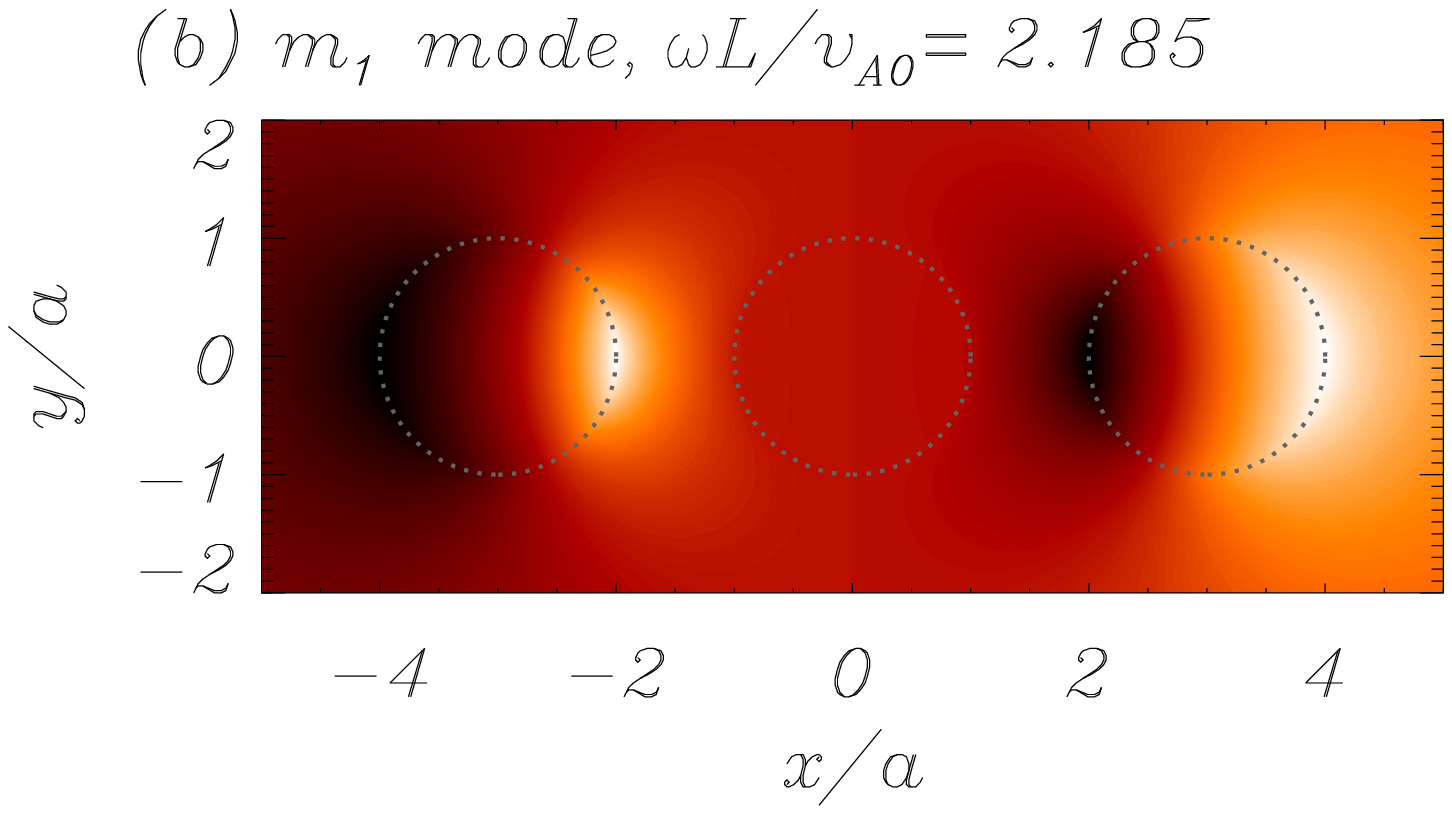}
\mbox{\hspace{1.cm}\hspace{4.5cm}\hspace{4.5cm}\hspace{1.cm}}\vspace{-2.4cm}
\includegraphics[width=8.5cm]{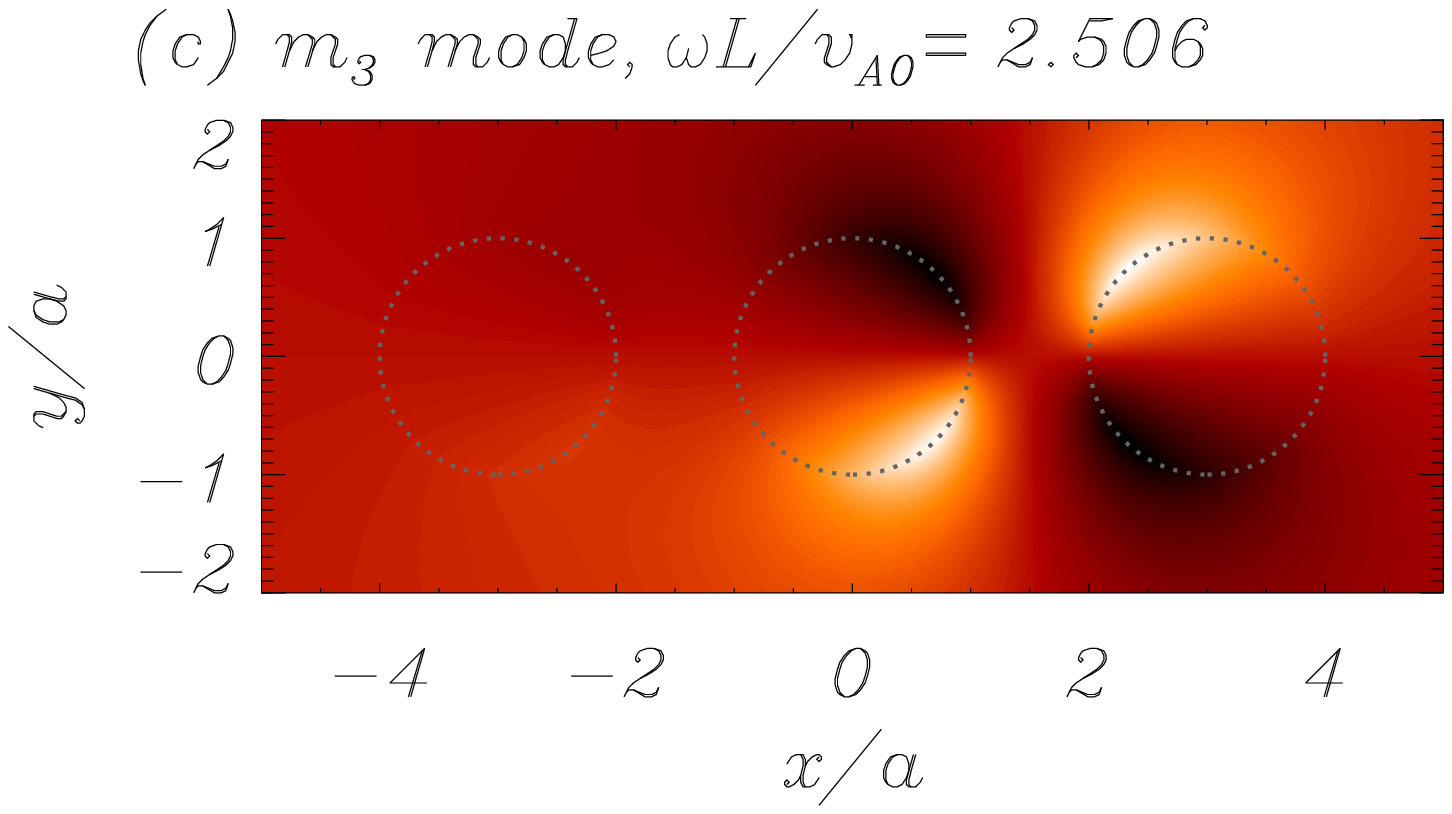}\hspace{-1cm}\includegraphics[width=8.5cm]{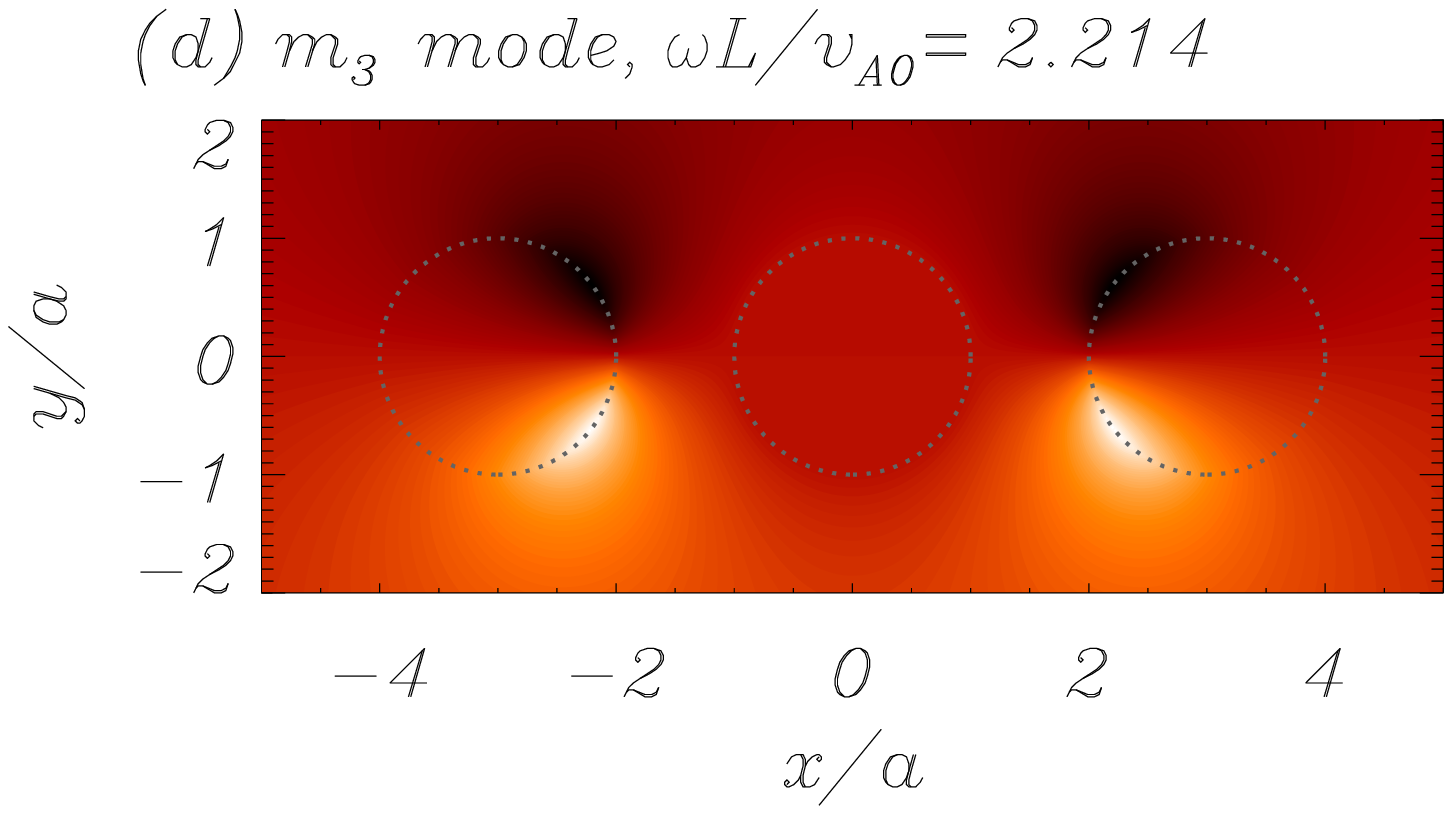}
\mbox{\hspace{1.cm}\hspace{4.5cm}\hspace{4.5cm}\hspace{1.cm}}\vspace{-2.4cm}
\includegraphics[width=8.5cm]{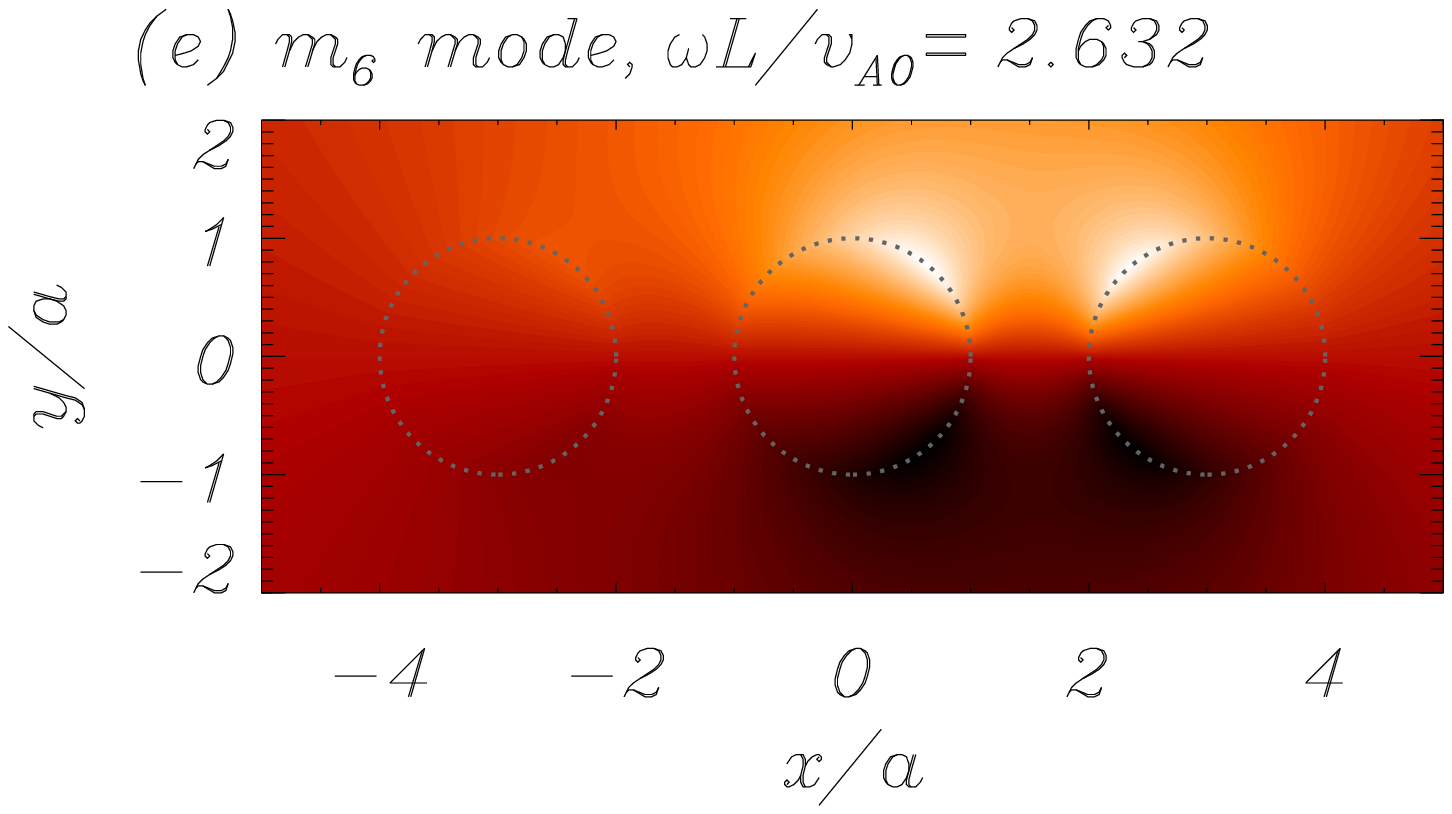}\hspace{-1cm}\includegraphics[width=8.5cm]{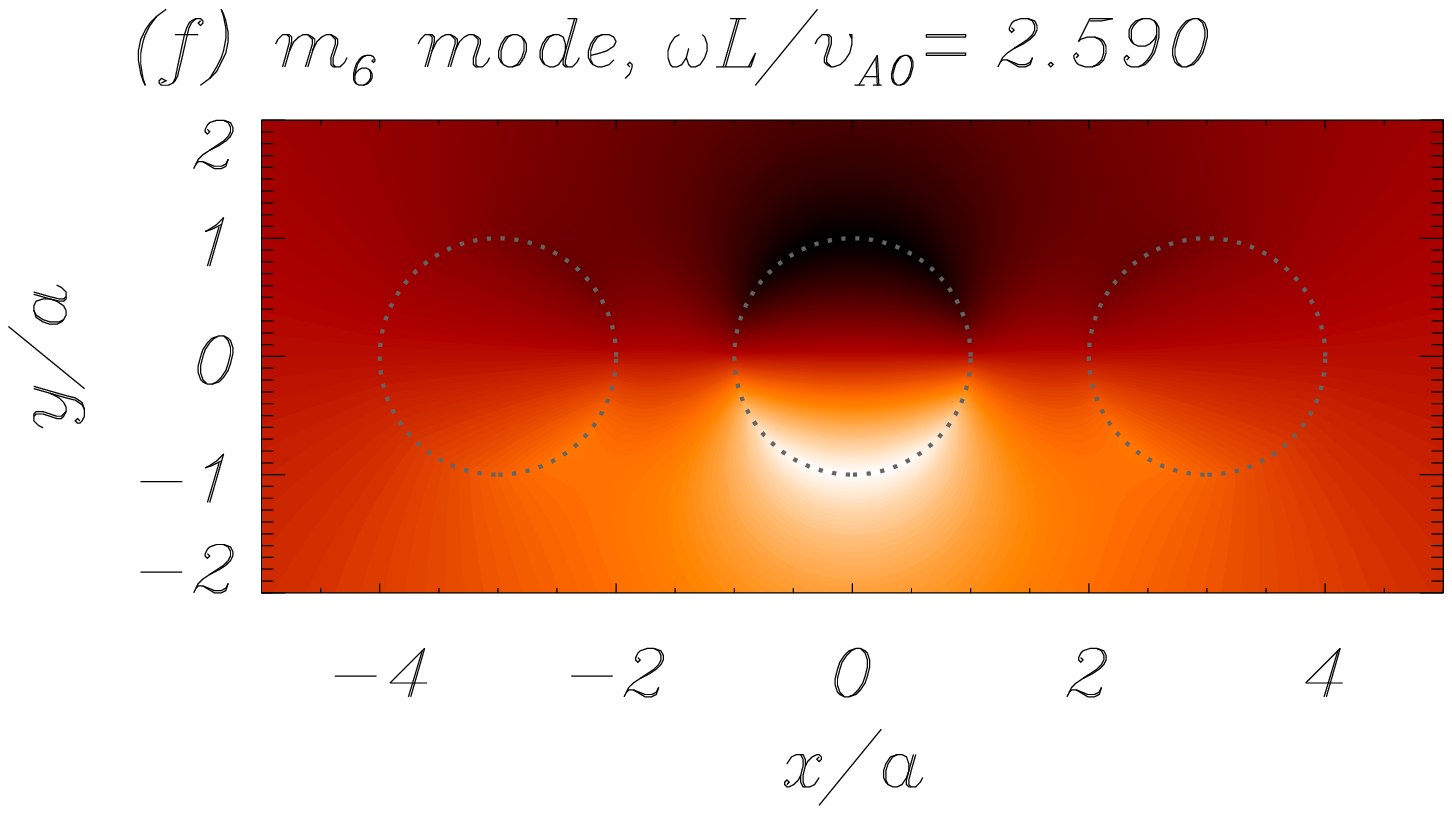}
\caption{Same as Fig. \ref{3loop_modes} for two values of $\rho_\mathrm{3}$ at
the maximal coupling (Fig. \ref{3loops_densities}). In the top, central, and
bottom rows of panels the $m_\mathrm{1}$, $m_\mathrm{3}$, and $m_\mathrm{6}$
modes are plotted for {\bf (a)}, {\bf (c)}, and  {\bf (e)} $\rho_\mathrm{3}=2
\rho_\mathrm{0}$; {\bf (b)}, {\bf (d)}, and  {\bf (f)} $\rho_\mathrm{3}=3
\rho_\mathrm{0}$.}
\label{3nonidloops_densities_coupled}
\end{figure}

Now we consider the dependence of the interaction on the loop density. The loop
radii are fixed to $a_\mathrm{1}=a_\mathrm{2}=a_\mathrm{3}=a=0.03 L$, the
separation between adjacent loops is $d=3 a$, the densities of loops $1$ and $2$
are fixed to $\rho_\mathrm{1}=3 \rho_\mathrm{0}$ and $\rho_\mathrm{2}=2
\rho_\mathrm{0}$, and $\rho_\mathrm{3}$ is changed from $\rho_\mathrm{0}$ to $4
\rho_\mathrm{0}$. Six kinklike normal modes, rather than eight, are found and
their frequencies are plotted as a function of $\rho_\mathrm{3}$ in Figure
\ref{3loops_densities}. There are six branches associated to the six kinklike
modes, that have been labeled $m_\mathrm{1}$ to $m_\mathrm{6}$ starting with the
lowest frequency mode. We have chosen $\rho_\mathrm{1}$ and $\rho_\mathrm{2}$ in
such a way that loops $1$ and $2$ are basically decoupled (see \S \ref{2loops}).
Figure \ref{3loops_densities} is similar to Figure \ref{ratio_densities} and can
be interpreted as two avoided crossings of the individual kink modes of the
three loops. Far from the couplings, the loops behave independently. This fact
is illustrated in Figure \ref{3nonidloops_densities_noncoupled}. In this figure
we have plotted the modes associated to the branches $m_\mathrm{1}$,
$m_\mathrm{3}$ and $m_\mathrm{6}$ in the top, central, and bottom rows,
respectively. The modes $m_\mathrm{2}$, $m_\mathrm{4}$ and $m_\mathrm{5}$ have
an equivalent behavior, and have not been plotted. Far from the coupling region
the $m_\mathrm{1}$ and $m_\mathrm{2}$ solutions are associated to the individual
kink oscillations of the denser loop in the $x$- and $y$-direction,
respectively. In the same way, the branches $m_\mathrm{3}$ and $m_\mathrm{4}$
are associated to the individual kink mode of the intermediate density loop and
the branches $m_\mathrm{5}$ and $m_\mathrm{6}$ to the individual kink
oscillations of the rarest loop. On the other hand, at the couplings the loops
interact by pairs as we see in Figure \ref{3nonidloops_densities_coupled}. The
interacting pair oscillates with a collective normal mode whereas the other loop
oscillates individually. In the first avoided crossing, for
$\rho_\mathrm{3}=\rho_\mathrm{2}=2 \rho_\mathrm{0}$, the branches $m_\mathrm{3}$
and $m_\mathrm{4}$ are coupled with $m_\mathrm{5}$ and $m_\mathrm{6}$ (see
Fig.~\ref{3loops_densities}), associated to loops $2$ and $3$, that oscillate
collectively as we see in Figures \ref{3nonidloops_densities_coupled}c and
\ref{3nonidloops_densities_coupled}e. The branches $m_\mathrm{1}$ and
$m_\mathrm{2}$ are uncoupled and loop $1$ oscillates independently from the
other two, as we see in Figure \ref{3nonidloops_densities_coupled}a. In the
second avoided crossing at $\rho_\mathrm{3}=\rho_\mathrm{1}=3 \rho_\mathrm{0}$
the branches $m_\mathrm{1}$ and $m_\mathrm{2}$ are coupled with $m_\mathrm{3}$
and $m_\mathrm{4}$, while $m_\mathrm{5}$ and $m_\mathrm{6}$ are independent.
Therefore, in this avoided crossing the interaction is between loops $1$ and $3$
(Figs.~\ref{3nonidloops_densities_coupled}b and
\ref{3nonidloops_densities_coupled}d) and loop $2$ oscillates independently
(Fig.~\ref{3nonidloops_densities_coupled}f). It is important to note that the
collective modes of the two coupled tubes have a different frequency ordering
with respect to the case of two loops, studied in \S \ref{2loops}, because of
the presence of loop $1$. 

Comparing Figures \ref{ratio_densities} and \ref{3loops_densities} the coupling
regions occur in a narrower range of density values in the three loop system
than for two tubes. The physical meaning is that only loops with similar
densities are coupled in the three loop ensemble. In this system, it is
important to note that in the second avoided crossing at $\rho_\mathrm{3}\approx
\rho_\mathrm{1}$ loop $2$ does not participate of the collective dynamics
despite being the closest tube to the interacting loops.

The results discussed so far in this subsection correspond to different
densities of loops $1$ and $2$. Nevertheless, if the densities of loops $1$ and
$2$ are similar, their interaction is more important and the description of the
dispersion diagram and the normal modes of the system is much more complex. In
this case, there are eight kinklike normal modes. In this situation there are
not only interactions between pairs of loops but also interactions between three
loops. There are modes associated to the ensemble formed by tubes $1$ and $2$,
individual oscillations of the cylinder $3$ and the ensemble of the three loops
depending on $\rho_\mathrm{3}$. A particular case of this situation is the three
identical loops previously discussed (\S \ref{eq_loops}) in which all modes
are associated to the collectivity.

\section{Discussion and conclusions}\label{disc_conc}

In this work we have investigated the kinklike normal modes of a system of
several loops with the help of the T-matrix theory. The results of this work can
be summarized as follows:

\begin{enumerate}

\item In the system of two non-identical loops, we have found four kinklike
normal modes $P_x$, $AP_y$, $P_y$, and $AP_x$. The frequencies of the $P_x$ and
$AP_y$ solutions are very similar as well as the frequencies of the $P_y$ and
$AP_x$ modes. This result agrees with \citet{doorsselaere2008}, who considered
thin tubes (i.e. long wavelength approximation). For fat loops the $P_x$ and
$AP_y$ modes, as well the $P_y$ and $AP_x$, have different frequencies, as was
shown in \cite{luna2008}.

\item For a system of two loops we have investigated the dependence of the
interaction between kink oscillations as a function on the relative density of
the loop pair. For $\rho_\mathrm{1}=3 \rho_\mathrm{0}$ we have found that the
oscillations of the loops are coupled in the range of $\rho_\mathrm{2}$ between
$2 \rho_\mathrm{0}$ to $4 \rho_\mathrm{0}$ and that the coupling is maximum at
$\rho_\mathrm{2} = \rho_\mathrm{1}=3\rho_\mathrm{0}$. Outside this density range
the loops are essentially decoupled and oscillate independently. This is
qualitatively similar to the behavior of the anomalous modes described by
\citet{doorsselaere2008}.

\item We have also studied the dependence of the interaction with the relative
radii of the loops. We have seen that in the range of radii for which transverse
loop oscillations have been observed the interaction depends very little on this
parameter and the loops strongly interact for all the radii considered. The explanation of this behavior is that in our loops the thin
loop approximation can be applied and in this situation the kink frequency
depends on the tube density and not on the radii.

\item In the case of a system of three equal, aligned loops there are eight
kinklike normal modes. The lower frequency mode corresponds to the three loops
oscillating in phase in the $x$-direction, i.e. along the direction in which
their axes are aligned, in agreement with the results of two identical loops. On
the other hand, the upper frequency mode corresponds to the three loops
oscillating in phase in the $y$-direction. This does not agree with the two
identical loops situation, in which the upper mode corresponds to the two loops
oscillating in antiphase in the $x$-direction. In fact, this property of the
three-loop system is also true for ensembles of four or more aligned loops.

\item We have made a parametric study of the kinklike modes in a system of three
loops with equal radii and different densities by changing the density of loop
$3$, $\rho_\mathrm{3}$. We have chosen $\rho_\mathrm{1}=3 \rho_\mathrm{0}$ and
$\rho_\mathrm{2}=2 \rho_\mathrm{0}$ so that the interaction between loops $1$
and $2$ is negligible. We have found that the oscillations of loop $3$ are
coupled with loop $2$ when $\rho_\mathrm{3} \approx \rho_\mathrm{2}$, whereas
loop $1$ oscillates independently. Furthermore, loop $3$ couples with loop $1$
when $\rho_\mathrm{3} \approx \rho_\mathrm{1}$ with loop $2$ oscillating
independently. If $\rho_\mathrm{3}$ takes different values, the system is
decoupled and the three loops oscillate independently.

\end{enumerate}

In this work, we have found that the interaction between loops regarding
kinklike motions depends strongly on the their individual kink frequencies. If
these frequencies are similar, loop motions are coupled and the normal modes are
collective. On the other hand, if the loop kink frequencies are quite different
their motions are not coupled. Since the individual frequencies depend on the
loop density and radius, we have studied separately the influence of the two
parameters. We have found that if the densities are quite similar, loops are
coupled and the oscillations are collective. On the other hand, if the densities
are quite different, the tubes oscillate independently. The range of densities
for which the loops are coupled depends on the system properties and in the
configuration of three loops this range is narrower than in the two tubes
configuration.

From the results shown in this paper we suggest that the antiphase motions
reported in \citet{schrijver2000} and \citet{schrijver2002} are collective
motions and, therefore, that the individual kink frequencies are similar but
different from the collective observed frequency. If the loop model presented
here is valid, both loop densities are also similar. In addition, in
\citet{verwichte2004} a loop arcade is studied and three groups of tubes
oscillating with similar frequencies can be appreciated. The dynamics of each
group of tubes can be interpreted as collective, although a detailed study of
such configuration is needed to relate the loop characteristics and the
frequency of oscillation of the group. On the other hand, loops not belonging to
these three groups do not share their frequencies with other loops, and so
oscillate independently. This has to be interpreted as a sign that these loops
have different densities from those of the rest of the loops. It must be
mentioned that in \citet{verwichte2004} all the oscillations are assumed as
individual, but this is only true in the case of loops that do not share their
frequency. If this assumption is applied to loop with a collective behavior it
produces wrong results for the loop parameters. For example, if the loops
actually oscillate with the lowest frequency collective mode, the assumption of
these authors produces an underestimation of the magnetic field or an
overestimation of the loop density.

The T-matrix method shown in this paper can be easily applied to more complex
configurations with gas pressure and tubes with flows or more complex systems of
loops, i.e. arcades with myriads of loops or multistranded loops. It is expected
that in such systems loops oscillate essentially independently except for loops
with similar individual oscillation frequencies.

\acknowledgments 

We would like to thank Dr. Rony Keppens for calling our attention about the
T-matrix theory and Dr. A. J. D\'{\i}az for his comments. M. Luna is grateful to
the Spanish Ministry of Science and Education for an FPI fellowship, which is
partially supported by the European Social Fund. He also thanks the members of
the Departament of Mathematics of K. U. Leuven for their warm hospitality during
his brief stay at this University and for their worthy comments. The authors
acknowledge the Spanish Ministry of  Science and Education and the Conselleria
d'Economia, Hisenda i Innovaci\'o of the Goverment of the Balearic Islands for
the funding provided under grants AYA2006-07637, PRIB-2004-10145, and
PCTIB-2005-GC3-03, respectively.

\end{document}